\documentclass[prd,preprint,nofootinbib,floats,superscriptaddress,tightenlines,amsfonts,amsmath,amssymb]{revtex4}

\usepackage{bm}
\usepackage{color}
\usepackage{graphicx}
\usepackage{multirow}
\usepackage{slashed}

\usepackage[colorlinks,linkcolor=black,anchorcolor=black,citecolor=black]{hyperref}

\begin{document}
\baselineskip=3ex \parskip=1ex

\preprint{NCTS-PH/1713}

\title{\boldmath$R_{K^{(*)}}$ and related $b\to s\ell\bar\ell$ anomalies in\\
minimal flavor violation framework with $Z'$ boson}

\author{Cheng-Wei Chiang}
\email[e-mail: ]{chengwei@phys.ntu.edu.tw}
\affiliation{Department of Physics, National Taiwan University, Taipei 10617, Taiwan}
\affiliation{Institute of Physics, Academia Sinica, Taipei 11529, Taiwan}
\affiliation{Physics Division, National Center for Theoretical Sciences, Hsinchu 30013, Taiwan}
\affiliation{Kavli IPMU, University of Tokyo, Kashiwa, 277-8583, Japan}

\author{Xiao-Gang He}
\email[e-mail: ]{hexg@phys.ntu.edu.tw}
\affiliation{Department of Physics, National Taiwan University, Taipei 10617, Taiwan}
\affiliation{Physics Division, National Center for Theoretical Sciences, Hsinchu 30013, Taiwan}
\affiliation{T-D Lee Institute, School of Physics and Astronomy, Shanghai Jiao Tong University, \\
Shanghai 200240, China \vspace*{3ex}}

\author{Jusak Tandean}
\email[e-mail: ]{jtandean@yahoo.com}
\affiliation{Department of Physics, National Taiwan University, Taipei 10617, Taiwan}
\affiliation{Physics Division, National Center for Theoretical Sciences, Hsinchu 30013, Taiwan}

\author{Xing-Bo Yuan}
\email[e-mail: ]{xbyuan@cts.nthu.edu.tw}
\affiliation{Physics Division, National Center for Theoretical Sciences, Hsinchu 30013, Taiwan}


\begin{abstract}
A recent LHCb measurement of the ratio $R_{K^*}$ of $B\to K^*\mu\bar\mu$ to $B\to K^*e\bar e$
branching fractions has produced results in mild tension with the standard model (SM).
This adds to the known anomalies also induced by the $b\to s\ell\bar\ell$ transitions, resulting
in a confidence level now as high as 4$\sigma$.
We analyze whether the parameter space preferred by all the $b\to s\ell\bar\ell$ anomalies is
compatible with a\,\,heavy $Z'$ boson assumed to have nonuniversal couplings to SM fermions
dictated by the principle of minimal flavor violation (MFV).
We deal with the MFV couplings of the $Z'$ to leptons in the context of the type-I seesaw
scenario for generating neutrino masses.
The flavor-violating $Z'$ interactions are subject to stringent constraints from other processes,
especially $B$-$\bar B$ mixing, charged lepton decays $\ell_i\to\ell_j\ell_k\bar\ell_l$ occurring
at tree level, and the loop induced $\mu\to e\gamma$.
We perform scans for parameter regions allowed by various data and predict the ranges for a number
of observables.
Some of the predictions, such as the branching fractions of lepton-flavor violating $\tau\to3\mu$,
$B\to K e\mu$, $K_L\to e\mu$, and $Z\to\ell\ell'$, are not far below their experimental bounds and
therefore could be probed by searches in the near future.
The viable parameter space depends strongly on the neutrino mass hierarchy, with a preference for
the inverted one.
\end{abstract}

\maketitle

\section{Introduction \label{sec:intro}}

In addition to direct searches for new physics (NP) at the energy frontier, the CERN LHC has
been testing the standard model (SM) of particle physics through studies of flavor physics.
While up to date there is still no strong evidence of nonstandard particles or interactions
predicted by various NP models, LHC experiments have, however, turned up quite a few anomalous
results in the lower energy regime.
In particular, a pattern of discrepancies from SM expectations has recently been emerging from
observables in a number of $b\to s\ell^+\ell^-$ transitions, mostly at around or above
the $3\sigma$ level.
Such coherent deviations call for special attention, as the observables are sensitive to
contributions from new particles and/or new interactions.

The aforementioned indications of anomalous \,$b\to s\ell^+\ell^-$\, interaction showed up in
the binned angular distribution of the \,$B\to K^*\mu^+\mu^-$\, decay, first found by the LHCb
Collaboration~\cite{Aaij:2013qta,Aaij:2015oid} and later on confirmed by the Belle
Collaboration~\cite{Abdesselam:2016llu,Wehle:2016yoi}.
The anomalies also include the observed deficits in the branching fractions of
\,$B\to K^{(*)}\mu^+\mu^-$ and $B_s\to\phi\mu^+\mu^-$\,
decays\,\,\cite{Aaij:2014pli,Aaij:2013aln,Aaij:2015esa,pdg}.
Another set of observables that have manifested unexpected values are
\begin{align}
\begin{split}
R_K &\equiv \frac{{\cal B}(B\to K \mu^+\mu^-)}{{\cal B}(B\to K e^+e^-)} ~,
\\
R_{K^*} &\equiv \frac{{\cal B}(B\to K^* \mu^+\mu^-)}{{\cal B}(B\to K^* e^+e^-)} ~,
\end{split}
\end{align}
first proposed in Ref.\,\cite{Hiller:2003js}.
These are of great interest because most of the hadronic uncertainties cancel out in the ratios,
and so they provide a sensitive test of lepton-flavor universality (LFU).
In the SM both $R_K$ and $R_{K^*}$ are predicted to be very close to
unity \cite{Hiller:2003js,Bouchard:2013mia,Bordone:2016gaq}.
However, the former was determined by LHCb~\cite{Aaij:2014ora} to be
\,$R_K=0.745_{-0.074}^{+0.090}{\rm(stat)}\pm 0.036{\rm(syst)}$\, for the dilepton invariant mass
squared range \,$q^2\in(1,6)\rm GeV^2$.\,
This finding can be reconciled with the corresponding SM value at the 2.6$\sigma$
level \cite{Aaij:2014ora}.
Very recently, LHCb \cite{Aaij:2017vbb} also reported a measurement on $R_{K^*}$:
\begin{eqnarray}
R_{K^*} \,=\, \begin{cases} 0.66^{+0.11}_{-0.07}{\rm(stat)}\pm 0.03{\rm(syst)}\qquad\mbox{for}~~
q^2\in(0.045,1.1)\,\mbox{GeV}^2 \,,
\smallskip \\
0.69^{+0.11}_{-0.07}{\rm(stat)}\pm 0.05{\rm(syst)}\qquad\mbox{for}~~q^2\in(1.1,6)\,\mbox{GeV}^2 \,.
\end{cases}
\end{eqnarray}
These are compatible with their SM counterparts \,$R_{K^*}^{\rm SM}=0.906(28)$ and 1.00(1)
\cite{Bordone:2016gaq}, respectively, at the 2.1$\sigma$ and 2.4$\sigma$
levels\,\,\cite{Aaij:2017vbb}.
The data on $R_K$ and $R_{K^*}$ together reveal consistent breaking of LFU at an even higher
confidence level (CL) of about $4\sigma$~\cite{DAmico:2017mtc}.
This has added to the tantalizing tentative hints of the presence of NP in these processes
which has the feature of violating LFU.
Thus, unsurprisingly the new $R_{K^*}$ anomaly has stimulated a new wave of theoretical studies
about lepton-flavor-nonuniversal \,$b\to s\ell^+\ell^-$\,
interactions~\cite{DAmico:2017mtc,Altmannshofer:2017yso,Capdevila:2017bsm,Hiller:2017bzc,
Geng:2017svp,Ciuchini:2017mik,Celis:2017doq,Becirevic:2017jtw,Cai:2017wry,Greljo:2017vvb,
Feruglio:2017rjo,Megias:2017isd,Hurth:2017hxg,Das:2017kfo,Dinh:2017smk,Bardhan:2017xcc,
Bordone:2017lsy,Neshatpour:2017qvi,Ko:2017yrd,Kamenik:2017tnu,DiChiara:2017cjq,Ghosh:2017ber,
Alok:2017jaf,Alok:2017sui,Wang:2017mrd,Alonso:2017bff,Bonilla:2017lsq,Ellis:2017nrp,
Bishara:2017pje,Alonso:2017uky,Tang:2017gkz,Datta:2017ezo,Matsuzaki:2017bpp,DiLuzio:2017chi}.
In this paper, we also entertain the possibility that these anomalies arise from LFU-violating
NP and explore some of its implications.

When addressing flavor physics beyond the SM, the usual problem one faces is that there are too
many model-dependent parameters.
On one hand, this provides an opportunity of having rich phenomenology in the flavor sector.
On the other hand, the sizable number of parameters tends to complicate the analysis, in some
cases making the situation arbitrary.
If there is a way to treat the flavor structure systematically, it may simplify the analysis
and provide a guide for theoretically understanding the potential NP.
One of the efficient means to this end is the framework of so-called minimal flavor violation
(MFV), which we will adopt.
The MFV principle postulates that Yukawa couplings are the sources of all flavor and $CP$
violations~\cite{mfv1,D'Ambrosio:2002ex}.
Applying the MFV idea to an effective field theory approach at low energies would then offer
a natural model-independent solution for TeV-scale NP to evade flavor-changing neutral current
(FCNC) restrictions.
Although initially motivated by the successful SM description of quark FCNCs, the notion of MFV
can be extended to the lepton sector~\cite{Cirigliano:2005ck}.
However, as the SM strictly does not accommodate lepton-flavor violation and it remains unknown
whether neutrinos are Dirac or Majorana particles, there is currently no unique way to implement
MFV in the lepton sector.
To do so will usually involve picking a particular scenario for endowing neutrinos with mass.

Our interest here is in studying within the MFV framework whether the parameter space preferred
by all the $b\to s\ell^+\ell^-$ anomalies have any conflict with other related observables.
After revisiting the case of the relevant dimension-six operators satisfying the MFV criterion in
both their quark and lepton parts, we will focus on a scenario in which the flavor violations are
induced by an electrically neutral and uncolored vector particle, such as a $Z'$ boson,
which has effective fermionic interactions consistent with the MFV principle.
We will look at a variety of constraints on its couplings to quarks and leptons and subsequently
evaluate a number of predictions from the allowed parameter space associated with this particle.

Now, recent global analyses~\cite{DAmico:2017mtc,Altmannshofer:2017yso,Capdevila:2017bsm,
Altmannshofer:2017fio} have demonstrated that the dimension-6 operators that can produce some
of the best fits to the anomalous $b\to s\ell^+\ell^-$ findings are given by
\begin{eqnarray} \label{Lsmnp}
{\mathcal L}_{\rm eff}^{} &\supset& \sqrt8\,G_{\rm F}^{}V_{ts}^*V_{tb}^{}\,
\big(C_9^\ell O_9^\ell+C_{10}^\ell O_{10}^\ell\big) \;+\; {\rm H.c.} \,,
\nonumber \\
O_9^\ell &=& \frac{\alpha_{\rm e}^{}}{4\pi}\,\bar s\gamma^\eta P_L^{}b\,
\overline{\ell}\gamma_\eta^{}\ell \,, ~~~~~
O_{10}^\ell \,=\, \frac{\alpha_{\rm e}^{}}{4\pi}\,\bar s\gamma^\eta P_L^{}b\,
\overline{\ell}\gamma_\eta^{}\gamma_5^{}\ell ~,
\end{eqnarray}
with \,$C_i^\ell=C_i^{\rm SM}+C_i^{\ell,\rm NP}$ ($i=9,10$) being Wilson coefficients and
the NP entering mainly the $\ell=\mu$ terms.
In these formulas, $G_{\rm F}$ is the Fermi decay constant, $\alpha_{\rm e}=1/133$ denotes
the fine structure constant at the $b$-quark mass ($m_b$) scale, $V_{ts,tb}$ are elements of
the Cabibbo-Kobayashi-Maskawa (CKM) mixing matrix $V_{\textsc{ckm}}$, at the $m_b$ scale
\,$C_9^{\rm SM}\simeq-C_{10}^{\rm SM}\simeq4.2$\, universally for all charged leptons,
and \,$P_L^{}=(1-\gamma_5)/2$.\,
Unlike\,\,$O_{9,10}^\ell$, dimension-6 quark-lepton operators with scalar or tensor structures
are not favored by the data~\cite{Altmannshofer:2017yso,Altmannshofer:2017fio,Alonso:2014csa}.
As will be seen below, the dimension-6 operators with MFV considered in this work generate
interactions that are chiral and feature the relation \,$C_9^{\ell,\rm NP}=-C^{\ell,\rm NP}_{10}$.\,
With the NP effect on the electron channel taken to be vanishing, the $1\sigma$ allowed range of
$C_9^{\mu,\rm NP}$ has been found to be $[-0.81,-0.48]$ in this scenario~\cite{Altmannshofer:2017yso}.
Assuming that the new interactions in the MFV framework are mediated by a putative $Z'$ gauge boson,
we will examine whether the implied parameter space is consistent with existing data on processes
such as $B$-$\bar B$ mixing, neutrino oscillations, and lepton-flavor violating (LFV) processes.

The paper is arranged as follows.  Section~\ref{sec:MFV} briefly reviews the idea of MFV and
explains what type of dimension-6 operators with MFV are compatible with the $b\to s\ell^+\ell^-$
anomalies.
In Section~\ref{sec:zpr}, we introduce a $Z'$ gauge boson that can effectively induce the desired
flavor-changing interactions.
Subsequently, we discuss how they can account for the \,$b\to s\ell^+\ell^-$\, anomalies and must
respect various constraints, especially from measurements of $B$-$\bar B$ and neutrino
oscillations and search bounds on LFV processes.
In Section~\ref{sec:analysis}, we scan the parameter space subject to these requirements and
illustrate the viable regions.
Among the restraints, we find that the decays \,$\mu\to e\gamma$\, and \,$\tau\to3\mu$\, may play
the most constraining role, depending on the ordering of light neutrinos' masses.
Section~\ref{sec:predictions} is dedicated to our predictions for a number of processes based upon
our parameter scan results.
Section~\ref{sec:summary} summarizes our findings.
An Appendix contains some extra information.

\section{Operators with minimal flavor violation \label{sec:MFV}}

Since the quark masses and mixing angles are now well determined, the application of
MFV in the quark sector is straightforward.
In contrast, there is no unique way to formulate leptonic MFV because our knowledge about
the nature and absolute scale of neutrino masses is far from complete.
Given that flavor mixing among neutrinos has been empirically established\,\,\cite{pdg},
it is attractive to implement leptonic MFV by integrating new ingredients that can
account for this fact~\cite{Cirigliano:2005ck}.
One could consider a minimal field content where only the SM fermionic doublets and singlets
transform nontrivially under the flavor group, with lepton number violation and neutrino
masses being ascribed to the dimension-five Weinberg operator~\cite{Cirigliano:2005ck}.
Less minimally, one could explicitly introduce right-handed neutrinos~\cite{Cirigliano:2005ck},
or alternatively right-handed weak-SU(2)-triplet fermions~\cite{He:2014efa},
which transform  nontrivially under an expanded flavor group and are responsible for
the seesaw mechanism giving Majorana masses to light neutrinos\,\,\cite{seesaw1,seesaw3}.
One could also introduce instead a weak-SU(2)-triplet of unflavored
scalars~\cite{He:2014efa,Gavela:2009cd} which take part in the seesaw
mechanism~\cite{seesaw2}.\footnote{Some other aspects or possibilities of leptonic MFV have
been discussed in the literature~\cite{Davidson:2006bd,Branco:2006hz,Joshipura:2009gi,
Alonso:2011jd,Sierra:2012yy,He:2014fva,He:2014uya,Alonso:2015sja,Pilaftsis}.}
Here we apply MFV to leptons by invoking the type-I seesaw scenario involving
three heavy right-handed neutrinos.

The renormalizable Lagrangian for the masses of SM fermions plus the right-handed neutrinos,
denoted by $N_{1,2,3}$, can be expressed as
\begin{eqnarray}
{\cal L}_{\rm m}^{} &=& -(Y_u)_{jk}^{}\,\overline{Q_j^{}}P_R^{}U_k^{} \tilde H
- (Y_d)_{jk}^{}\,\overline{Q_j^{}}P_R^{}D_k^{}H - (Y_e)_{jk}^{}\,\overline{L_j^{}}P_R^{}E_k^{}H
\nonumber \\ && -\; (Y_\nu)_{jk}^{}\,\overline{L_j^{}}P_R^{}N_k^{}\tilde H \,-\,
\tfrac{1}{2}(M_N)_{jk}^{}\,\overline{\big(N_j^{}\big)\raisebox{2pt}{$^{\rm c}$}}P_R^{}N_k^{}
\;+\; {\rm H.c.} \,,
\end{eqnarray}
where summation over the generation indices \,$j,k=1,2,3$\, is implicit, $Y_{u,d,e,\nu}$ are
Yukawa coupling matrices, the quark, lepton, and Higgs doublets are given by
\begin{eqnarray}
Q_k^{} \,= \left(\begin{array}{c} U_k^{} \smallskip \\ D_k^{} \end{array}\right) , ~~~~ ~~~
L_k^{} \,=
\left(\begin{array}{c} \nu_k^{} \smallskip \\ E_k^{} \end{array}\right) , ~~~~ ~~~
H \,= \left(\begin{array}{c} 0  \smallskip \\ \tfrac{1}{\sqrt2}(h+v)
\end{array}\right) , ~~~~~
\tilde H \,=\, i\tau_2^{}H^*
\end{eqnarray}
after electroweak symmetry breaking, with \,$v\simeq246$ GeV\, being the vacuum expectation value
of $H$ and $\tau_2^{}$ the second Pauli matrix, $M_N$ is the Majorana mass matrix for $N_{1,2,3}$
which without loss of generality can be chosen to be diagonal, the superscript in $(N_j)^{\rm c}$
refers to charge conjugation, and \,$P_R^{}=(1+\gamma_5^{})/2$.\,
Hereafter, we entertain the possibility that $N_{1,2,3}$ are degenerate, and so
\,$M_N={\cal M}\,{\rm diag}(1,1,1)$.\,
It is then realized that ${\cal L}_{\rm m}$ is formally invariant under the global flavor
rotations \,$Q\rightarrow V_Q^{}Q$, \,$P_R^{}U\to P_R^{}V_U^{}U$, \,$P_R^{}D\to P_R^{}V_D^{}D$,
\,$L\to V_L^{}L$, \,$P_R^{}E\to P_R^{}V_E^{}E$,\, and
\,$N=(N_1\,\,\,N_2\,\,\,N_3)^{\textsc t}\to{\cal O}_N^{}N$,\,
with \,$V_{Q,U,D,L,E}\in{\rm SU}(3)_{Q,U,D,L,E}$\, and ${\cal O}_N$ being a real
orthogonal matrix, provided that the Yukawa couplings behave like spurions
transforming as
\,$Y_u^{}\to V_Q^{}Y_u^{}V_U^\dagger$, \,$Y_d^{}\to V_Q^{}Y_d^{}V_D^\dagger$,
\,$Y_e^{}\to V_L^{}Y_e^{}V_E^\dagger$,\, and
\,$Y_\nu^{}\to V_L^{}Y_\nu^{}{\cal O}_N^{\textsc t}$.\,

The right-handed neutrinos' mass, $\cal M$, is assumed to be very large compared to
the elements of $v Y_\nu/\sqrt2$, triggering the type-I seesaw mechanism~\cite{seesaw1}
which brings about the light-neutrinos' mass matrix
\,$m_\nu^{}=-(v^2/2)Y_\nu^{}M_N^{-1}Y_\nu^{\textsc t}=
U_{\textsc{pmns}\,}^{}\hat m_{\nu\,}^{}U_{\textsc{pmns}}^{\textsc t}$,\,
where $U_{\textsc{pmns}}$ is the Pontecorvo-Maki-Nakagawa-Sakata~\cite{pmns} mixing
matrix and \,$\hat m_\nu^{}={\rm diag}\bigl(m_1^{},m_2^{},m_3^{}\bigr)$\,
contains the light neutrinos' eigenmasses, $m_{1,2,3}$.
This suggests adopting the interesting form~\cite{Casas:2001sr}
\begin{eqnarray} \label{Ynu}
Y_\nu^{} \,=\,
\frac{i\sqrt2}{v}\,U_{\textsc{pmns}\,}^{}\hat m^{1/2}_\nu OM_N^{1/2} \,,
\end{eqnarray}
where $O$ is a generally complex orthogonal matrix satisfying
\,$OO^{\textsc t}=\openone\equiv{\rm diag}(1,1,1)$.\,

The MFV framework presupposes that the Yukawa couplings are the only sources of flavor and $CP$
violations~\cite{mfv1,D'Ambrosio:2002ex}.
Accordingly, to construct effective Lagrangians beyond the SM with MFV built-in, one inserts
products of the Yukawa matrices among the pertinent fields to devise operators that are
singlet under the SM gauge group and invariant under the flavor rotations described
above~\cite{D'Ambrosio:2002ex}.
Of potential interest here are the combinations
\begin{eqnarray} \label{AB}
\textsf{A}_q^{} \,=\, Y_u^{}Y_u^\dagger \,, ~~~~~
\textsf{B}_q^{} \,=\, Y_d^{}Y_d^\dagger \,, ~~~~~~~
\textsf{A}_\ell^{} \,=\, Y_\nu^{}Y_\nu^\dagger \,, ~~~~~
\textsf{B}_\ell^{} \,=\, Y_e^{}Y_e^\dagger \,.
\end{eqnarray}
With these, one assembles for the quark (lepton) sector an object $\Delta_q$ ($\Delta_\ell$) which,
in a model-independent approach, is formally an infinite series comprising all possible products of
$\textsf{A}_q$ and $\textsf{B}_q$ ($\textsf{A}_\ell$ and $\textsf{B}_\ell$).
The MFV hypothesis dictates that the series coefficients be real because otherwise they would
constitute new sources of $CP$ violation beyond the Yukawa couplings.
It turns out that, with the aid of the Cayley-Hamilton identity, one can resum the infinite series
in $\Delta_q$ $(\Delta_\ell)$ into a finite one consisting of merely seventeen
terms~\cite{Colangelo:2008qp}.
Because of the resummation, in this finite series the seventeen coefficients, denoted here by
$\zeta_r^{}$ $\big(\xi_r^{}\big)$ for \,$r=0,1,\ldots,16$,\, generally become complex.
However, it can be shown that
\,${\rm Im}\,\zeta_r^{}\propto\big|{\rm Tr}\big(\textsf{A}{}_q^2\textsf{B}_q^{}\textsf{A}_q^{}
\textsf{B}_q^2\big)\big|\ll1$ \cite{Colangelo:2008qp,He:2014uya}
and therefore these imaginary parts can be neglected in practical calculations.
The same can be said of the imaginary parts of $\xi_r^{}$.

Given that the maximum eigenvalues of $\textsf{A}_q$ and $\textsf{B}_q$ are, respectively,
\,$y_t^2=2m_t^2/v^2\simeq0.99$\, and \,$y_b^2=2m_b^2/v^2\simeq3.0\times10^{-4}$\, at the mass scale
\,$\mu=m_Z^{}$,\, for our purposes we can retain in $\Delta_q$ only terms up to two powers of
$\textsf{A}_q$ and drop terms with at least one power in $\textsf{B}_q$.
In $\Delta_q$, none of the 17 terms involves $\textsf{A}_q^3$ because it can be connected to
$\textsf{A}{}_q^{}$ and $\textsf{A}_q^2$ by means of the Cayley-Hamilton identity.
For the leptonic object $\Delta_\ell$, we select the right-handed neutrinos' mass $\cal M$
to be sufficiently large to make the maximum eigenvalue of $\textsf{A}_\ell$ equal unity.
Thus, as in the quark sector, we will keep in $\Delta_\ell$ only terms up to order
$\textsf{A}_\ell^2$ and ignore those with $\textsf{B}_\ell$, whose elements are at most
\,$y_\tau^2=2m_\tau^2/v^2\simeq1.0\times10^{-4}$.\,
It follows that the relevant spurion building blocks are
\begin{eqnarray} \label{DqDl}
\Delta_q^{} \,=\, \zeta_0^{}\openone + \zeta_1^{~}\textsf{A}_q^{}
+ \zeta_{2\,}^{}\textsf{A}_q^2 \,, ~~~~~~~
\Delta_\ell^{} \,=\, \xi_0^{}\openone + \xi_1^{~}\textsf{A}_\ell^{}
+ \xi_{2\,}^{}\textsf{A}_\ell^2 \,,
\end{eqnarray}
where, model-independently, the coefficients $\zeta_{0,1,2}^{}$ and $\xi_{0,1,2}^{}$ are free
parameters expected to be at most of ${\mathcal O}(1)$, with negligible imaginary
components~\cite{He:2014uya,Colangelo:2008qp}.
It is worth noting that these formulas are not the leading terms in expansions of the Yukawa
couplings, but the most general expressions for $\Delta_{q,\ell}$ after the $\textsf{B}_{q,\ell}$
contributions are neglected.\footnote{This appears to be in keeping with our findings later on.
Specifically, we obtain \,$\big|\zeta_1^{~}y_t^2+\zeta_2^{~}y_t^4\big|/m_{Z'}^{}<0.13$/TeV\, in
Sec. \ref{sec:4q} from the data on $B_s$-$\bar B_s$ mixing and
\,$\big|\zeta_0^{}\big|/m_{Z'}^{}\;\mbox{\footnotesize$\lesssim$}\;8\!\times\!10^{-6}$/TeV\, in 
Appendix\,\ref{2q2l} from the experimental bounds on \,$\mu\to e$\, conversion in nuclei.\medskip}
For the particular $Z'$-mediated interactions to be discussed in the next section, the nature of
the $Z'$ couplings to SM fermions implies that only the Hermitian portions of $\Delta_{q,\ell}$
matter.

It is convenient to work in the basis where $Y_{d,e}$ are diagonal,
\begin{eqnarray} \label{YdYe}
Y_d^{} \,=\, {\rm diag}\bigl(y_d^{},y_s^{},y_b^{}\bigr) \,, ~~~~ ~~~
Y_e^{} \,=\, {\rm diag}\bigl(y_e^{},y_\mu^{},y_\tau^{}\bigr) \,,
\end{eqnarray}
with \,$y_f^{}=\sqrt2\,m_f^{}/v$,\, and $U_k$, $D_k$, $\tilde\nu_{k,L}$, $N_{k,R}$,
and $E_k$ refer to the mass eigenstates.
In that case,
\begin{eqnarray} \label{AqAl}
Q_j^{} &=& \left(\!\begin{array}{c} \raisebox{1pt}{\footnotesize$\sum$}_k^{~}
\big(V_{\textsc{ckm}}^\dagger\big)_{jk} U_k^{} \vspace{2pt} \\ D_j^{} \end{array}\!\right) , ~~~~~~~
L_j^{} \,=\, \left(\!\begin{array}{c} \raisebox{1pt}{\footnotesize$\sum$}_k^{~}
(U_{\textsc{pmns}})_{jk\,}^{}\tilde\nu_k^{} \vspace{2pt} \\ E_j^{} \end{array}\!\right) , ~~~~~~~
Y_u^{} \,=\, V^\dagger_{\textsc{ckm}}\,{\rm diag}\bigl(y_u^{},y_c^{},y_t^{}\bigr) \,,
\nonumber \\
\textsf{A}_q^{} &=& V^\dagger_{\textsc{ckm}}\;
{\rm diag}\big(y_u^2,y_c^2,y_t^2\big)\, V_{\textsc{ckm}}^{} \,, ~~~~~~~
\textsf{A}_\ell^{} \,=\, \frac{2\mathcal M}{v^2}\, U_{\textsc{pmns}}^{}\, \hat m^{1/2}_\nu
O O^\dagger\hat m^{1/2}_\nu U_{\textsc{pmns}}^\dagger \,.
\end{eqnarray}
From this point on, we write \,$\ell_k=E_k$,\, and so \,$(\ell_1,\ell_2,\ell_3)=(e,\mu,\tau)$.

Without introducing other new interactions or particles, one then sees that the operators of
lowest dimension that are flavor invariant, SM gauge singlet, and of the type that can
readily give rise to the NP terms in Eq.\,(\ref{Lsmnp}) are~\cite{Lee:2015qra}
\begin{eqnarray}
O_1^6 &=& \overline{Q}\gamma_{\eta\,}^{}\Delta_{q\,}^{}P_L^{}Q\,
\overline{L}\gamma^\eta\Delta_{\ell\,}^{}P_L^{}L \,, ~~~~~~~
\nonumber \\
O_2^6 &=& \overline{Q}\gamma_{\eta\,}^{}\tilde\Delta_{q\,}^{}P_L^{}\tau_{a\,}^{}
Q\,\overline{L}\gamma^\eta\tilde\Delta_{\ell\,}^{}P_L^{}\tau_{a\,}^{}L \,,
\end{eqnarray}
where $\tilde\Delta_q$ and $\tilde\Delta_\ell$ are, respectively, of the same form as
$\Delta_q$ and $\Delta_\ell$ in Eq.\,(\ref{DqDl}), but have their own independent coefficients
$\tilde\zeta_r^{}$ and $\tilde\xi_r^{}$, and the index \,$a=1,2,3$\, of the Pauli matrix
$\tau_a^{}$ is implicitly summed over.
The MFV effective Lagrangian of interest is then
\begin{eqnarray} \label{Lmfv}
{\mathcal L}_{\rm eff}^{\textsc{mfv}} \,=\, \frac{1}{\Lambda^2}\big(O_1^6\,+\,O_2^6\big) \,,
\end{eqnarray}
where the mass scale $\Lambda$ characterizes the heavy NP underlying these interactions.

From Eq.\,(\ref{Lmfv}), one could obtain interactions that can account for
the $b\to s\ell^+\ell^-$ anomalies and investigate some of the implications~\cite{Lee:2015qra}
without explicitly addressing the underlying NP.
Specifically, among \,$b\to(s,d)\ell\bar\ell'$\, and \,$s\to d\ell\bar\ell'$\, decays with
\,$\ell\neq\ell'$\, as well as related processes with neutrinos in the final states, there
could be predicted rates which are not far from their experimental results and, therefore,
may be testable in near future searches~\cite{Lee:2015qra}.

In the rest of this paper, we concentrate instead on a scenario in which a $Z'$ gauge
boson with nonuniversal couplings to SM fermions is responsible for the NP effects
on \,$b\to s\ell^+\ell^-$.
Such a\,\,particle exists in many models~\cite{Langacker:2008yv}.\footnote{Recent literature with
regard to the \,$b\to s\ell\bar\ell$\, anomalies in the contexts of other models possessing
some kind of $Z'$ particle includes~\cite{Ko:2017yrd,Kamenik:2017tnu,DiChiara:2017cjq,
Ghosh:2017ber,Alok:2017jaf,Alok:2017sui,Wang:2017mrd,Alonso:2017bff,Bonilla:2017lsq,
Ellis:2017nrp,Bishara:2017pje,Alonso:2017uky,Tang:2017gkz,Datta:2017ezo,Matsuzaki:2017bpp,
DiLuzio:2017chi,Crivellin:2015lwa,Crivellin:2015era,Belanger:2015nma,Allanach:2015gkd,
Chiang:2016qov,Becirevic:2016zri,Kim:2016bdu,Altmannshofer:2016jzy,Bhattacharya:2016mcc}.}
Since $O_2^6$ contains charged-currents, only $O_1^6$ is attributable to the $Z'$
contribution at tree level.
It is worth remarking that, although this analysis concerns the $Z'$ gauge boson,
the main results are applicable to any new electrically neutral, uncolored, spin-1 particle,
which could be composite, having similar flavor-violating couplings.

\section{\boldmath$Z'$-Mediated Interactions \label{sec:zpr}}

The renormalizable Lagrangian for the interactions between SM fermions and the $Z'$ boson
fulfilling the MFV criterion can take the form \cite{Kim:2016bdu}
\begin{eqnarray} \label{LZ'}
{\cal L}_{Z'}^{} \,=\, -\big( \overline{Q}\gamma^\eta\Delta_q^{}P_L^{}Q
+ \overline{L}\gamma^\eta\Delta_\ell^{}P_L^{}L \big) Z'_\eta ~,
\end{eqnarray}
where any overall coupling constant of the $Z'$ has been absorbed into the coefficients
$\zeta_{0,1,2}^{}$ and $\xi_{0,1,2}^{}$ in $\Delta_q$ and $\Delta_\ell$, respectively,
as defined in Eq.\,(\ref{DqDl}).
These coefficients are now purely real because the Hermiticity of ${\cal L}_{Z'}$ implies
that $\Delta_{q,\ell}$ in Eq.\,\eqref{LZ'} are Hermitian as well.
We also suppose that any mixing between the $Z'$ and SM gauge bosons is negligible and that
the $Z'$ mass, $m_{Z'}^{}$, is above the electroweak scale.

From Eq.\,\eqref{LZ'}, one can readily derive the MFV Lagrangian, ${\mathcal L}_{\textsc{mfv}}$,
that involves three types of effective four-fermion operators with dimension up to 6.
Thus, besides $O^6_1$, the additional operators that can appear due to $Z'$ exchange at tree
level are given by
\begin{eqnarray} \label{Lmfv'}
{\mathcal L}_{\textsc{mfv}}^{} &=&
\frac{-1}{m_{Z'}^2} \big({\cal O}^{4q} + {\cal O}^{4\ell} + {\cal O}^{2q2\ell} \big) ~,
\\
{\cal O}^{4q} &=& \tfrac{1}{2}\,\overline{Q}\gamma_{\eta\,}^{}\Delta_q^{}P_L^{}Q\;
\overline{Q}\gamma^\eta\Delta_q^{}P_L^{}Q \,,
\nonumber
\\
{\cal O}^{4\ell} &=& \tfrac{1}{2}\,\overline{L}\gamma_{\eta\,}^{}\Delta_\ell^{}P_L^{}L\;
\overline{L}\gamma^\eta\Delta_\ell^{}P_L^{}L \,,
\nonumber
\\
{\cal O}^{2q2\ell} &=& O^6_1 \,=\,
\overline{Q}\gamma_{\eta\,}^{}\Delta_q^{}P_L^{}Q\;
\overline{L}\gamma^\eta\Delta_\ell^{}P_L^{}L \,,
\nonumber
\end{eqnarray}
where $m_{Z'}^{}$ is taken to be large compared to the energies of the external fermions.
With the extra operators to consider, we will need to deal with more constraints than
in a model-independent analysis based on the $\overline{Q}Q\overline{L}L$ operators in
Eq.\,(\ref{Lmfv}) alone.

In the following, we discuss the effects of ${\cal O}^{4q}$, ${\cal O}^{4\ell}$, and
${\cal O}^{2q2\ell}$ in turn and study the restrictions on the elements of $\Delta_{q,\ell}$
from existing data.
In view of the recent great interest in the \,$b\to s\ell^+\ell^-$\, anomalies, we start
with a discussion on the interactions involving ${\cal O}^{2q2\ell}$.

\subsection{Diquark-Dilepton Interactions \label{sec:2q2l}}

In the presence of ${\cal O}^{2q2\ell}$ in ${\mathcal L}_{\textsc{mfv}}$, the effective
interaction responsible for \,$b\to s\ell\bar\ell'$\, is
\begin{eqnarray} \label{b2qll'}
{\mathcal L}_{\rm eff}^{} &\supset& \frac{\sqrt2\,\alpha_{\rm e\,}^{}
\lambda_{sb\,}^{}G_{\rm F}^{}}{\pi}\, C_{\ell\ell'}^{}\,\overline{s}\,\gamma^\eta P_L^{}b\,
\overline{\ell}_{\,\!}\gamma_\eta^{}P_L^{}\ell' \,,
\end{eqnarray}
where
\begin{eqnarray} \label{Cll'}
\lambda_{q'q}^{} \,=\, V_{tq'}^*V_{tq}^{} \,, ~~~~~~~
C_{\ell\ell'}^{} \,=\, \delta_{\ell\ell'\,}^{} C_9^{\rm SM} \,+\, {\textsf c}_{\ell\ell'}^{} \,,
\end{eqnarray}
with the approximation \,$C_{10}^{\rm SM}=-C_9^{\rm SM}$.\,
Hence, in terms of the elements of $\Delta_{q,\ell}$
\begin{eqnarray} \label{cll'}
\textsf{c}_{\ell_j^{}\ell_k^{}}^{} \,=\, \frac{-\pi\,(\Delta_q)_{23\,}^{}(\Delta_\ell)_{jk}^{}}
{\sqrt2\,\alpha_{\rm e\,}^{}\lambda_{sb\,}^{}G_{\rm F\,}^{}m_{Z'}^2}
\,\simeq\, -25.3{\rm\,TeV}^2~
\frac{\big(\zeta_1^{~}y_t^2+\zeta_2^{~}y_t^4\big)(\Delta_\ell)_{jk}^{}}{m_{Z'}^2} \,,
\end{eqnarray}
where \,$(\Delta_q)_{23}^{}=\lambda_{sb}^{}\big(\zeta_1^{~}y_t^2 + \zeta_2^{~}y_t^4\big)$,\,
the contributions involving $y_{u,c}^{}$ having been dropped.
It follows that $|C_{\ell\ell'}|=|C_{\ell'\ell}|$.
Analogously, one can write down the corresponding expressions for $b\to d\ell\bar\ell'$
and $s\to d\ell\bar\ell'$.

Subsequent to the recent LHCb finding on $R_{K^*}$, it has been pointed out that one of the best
fits to the $b\to s\ell^+\ell^-$ data has the NP Wilson coefficients~\cite{Altmannshofer:2017yso}
\begin{eqnarray} \label{ceecmm}
\textsf{c}_{ee}^{} \,=\, 0 \,, ~~~~~~~
\mbox{$-1.00$} \,\le\, \textsf{c}_{\mu\mu}^{} \,\le\, -0.32
\end{eqnarray}
at the 2$\sigma$ level, which can be interpreted to imply that the $Z'$ boson does not couple
to electrons.
This is the scenario that we will continue to analyze in this work.
Since \,$\textsf{c}_{ee}^{}\propto(\Delta_\ell)_{11}^{}$,\, we then have from Eq.\,(\ref{ceecmm})
the condition \,$(\Delta_\ell)_{11}^{}=0$.\,

The same operator, ${\cal O}^{2q2\ell}$, contributes at tree level to \,$\mu\to e$\, conversion
in nuclei which is subject to stringent empirical limits.
Nevertheless, as outlined in Appendix\,\,\ref{2q2l}, the ${\cal O}^{2q2\ell}$ contribution to
this process can be made consistent with its current data by sufficiently reducing the size of
the coefficient $\zeta_0^{}$ in $\Delta_q$.

There may also be constraints from collider data.
However, given that \,$(\Delta_\ell)_{11}^{}=0$,\, limits implied by LEP
measurements on \,$e^+e^-\to q\bar q$\, \cite{Alcaraz:2006mx} can be evaded.
Moreover, our numerical calculations show that potential restraints from recent LHC results on
\,$pp\to\mu^+\mu^-$\, \cite{ATLAS:2017wce} are not yet realized,
as sketched in Appendix\,\,\ref{2q2l}.

\subsection{Four-Quark Interactions \label{sec:4q}}

The operator ${\cal O}^{4q}$ in ${\mathcal L}_{\textsc{mfv}}$ contributes at tree level to
the heavy-light mass difference of neutral $B_d$ ($B_s$) mesons, $\Delta M_{d(s)}$.
Including the SM contribution, we express it as~\cite{Buras:2012jb}
\begin{eqnarray}
\Delta M_{d(s)}^{} \,=\,
\Delta M_{d(s)}^{\textsc{SM}} \Bigg|1+\frac{S_{d(s)}^{Z'}}{S_0^{}(x_t^{})}\Bigg| \,,
\end{eqnarray}
where \,$S_0^{}(x_t)=2.35$\, for \,$m_t^{}=165$\,GeV\, is due to SM loop diagrams and
the $Z'$ part is
\begin{eqnarray}
S_{d(s)}^{Z'} \,=\,
\frac{4(\Delta_q)_{13(23)\,}^2\tilde r}{\lambda_{db(sb)}^2\,g_{\textsc{sm}}^2\,m_{Z'}^2} \,=\,
\frac{4\big(\zeta_1^{~}y_t^2+\zeta_2^{~}y_t^4\big)^2\tilde r}{g_{\textsc{sm}}^2\,m_{Z'}^2} \,,
\end{eqnarray}
with~\cite{Buras:2012jb} \,$g_{\textsc{sm}}^2=1.78\times10^{-7}\,\rm GeV^{-2}$\, and
the QCD factor \,$\tilde r\sim1$\, for \,$m_{Z'}^{}\sim1$\,TeV.\,

The experimental and SM values of $\Delta M_{d,s}$ are, in units of ps$^{-1}$,
\begin{eqnarray}
\Delta M_d^{\rm exp} &=& 0.5064\pm0.0019 ~~\mbox{\cite{hfag}} \,, ~~~~~~~
\Delta M_d^{\textsc{sm}} \,=\, 0.575_{-0.090}^{+0.093} ~~\mbox{\cite{Artuso:2015swg}} \,,
\nonumber \\
\Delta M_s^{\rm exp} &=& 17.757\pm0.021 ~~\mbox{\cite{hfag}} \,, ~~~~~ ~~~\,
\Delta M_s^{\textsc{sm}} \,=\, 18.6_{-2.3}^{+2.4} ~~\mbox{\cite{Artuso:2015swg}} \,,
\end{eqnarray}
where updated parameters have been used in the SM predictions.
From these numbers, we can calculate the 2$\sigma$ ranges
\begin{eqnarray}
0.60 \,\le\, C_{B_d}^{}=\frac{\Delta M_d^{\rm exp}}{\Delta M_d^{\textsc{sm}}} \,\le\, 1.16 \,,
~~~~~~~
0.71 \,\le\, C_{B_s}^{}=\frac{\Delta M_s^{\rm exp}}{\Delta M_s^{\textsc{sm}}} \,\le\, 1.19
\end{eqnarray}
after combining in quadrature the relative errors in the measurements and
predictions.
The first, and somewhat stronger, upper limit of these two constraints then translates
into\footnote{Employing instead the results \,$0.81\le C_{B_d}\le1.28$\, and
\,$0.899\le C_{B_s}\le1.252$,\, both at 95\% CL, of a global fit to constrain potential NP
contributions to \,$|\Delta F|=2$\, transitions~\cite{utfit}, reported at the end of last
summer, would yield a more relaxed condition than Eq.\,(\ref{constr}).}
\begin{eqnarray} \label{constr}
0 \,\le\, \frac{S_{d}^{Z'}}{S_0^{}(x_t^{})} \,=\, 9.56\times10^6{\rm\,GeV}^2~
\frac{\big(\zeta_1^{~}y_t^2+\zeta_2^{~}y_t^4\big)^2\tilde r}{m_{Z'}^2}
\,\le\, 0.16
\end{eqnarray}
or, with \,$\tilde r=1$,\,
\begin{eqnarray} \label{zyt}
\frac{\big|\zeta_1^{~}y_t^2+\zeta_2^{~}y_t^4\big|}{m_{Z'}^{}} \,\le\, \frac{0.13}{\rm TeV} \,.
\end{eqnarray}
This caps the quark part of $\textsf{c}_{\ell_j\ell_k}$ in Eq.\,(\ref{cll'}).

It is worth noting that the neutral-kaon system can furnish a comparable, but weaker,
restraint, as ${\cal O}^{4q}$ can modify the SM predictions for the $K_L$-$K_S$ mass
difference $\Delta M_K$ and the $CP$-violation parameter $\epsilon_K^{}$.
The $Z'$ contribution
\,$M_{12}^{K,Z'}=\big(V_{td}^*V_{ts}^{}\big)\raisebox{1pt}{$^2$}
\big(\zeta_1^{~}y_t^2+\zeta_2^{~}y_t^4\big)\raisebox{1pt}{$^{2\,}$}
\eta_{2\,}^{}\hat B_K^{}f_K^2m_{K^0\,} \tilde r/\big(6m_{Z'}^2\big)$,\,
with~\cite{Buras:2012jb} \,$\eta_2^{}=0.5765\pm0.0065$,\,
$\hat B_K=0.767\pm0.010$,\, and \,$f_K^{}=(156.1\pm1.1)$\,MeV,\, enters via
\,$\Delta M_K=2\,{\rm Re}\big(M_{12}^{K,\rm SM}+M_{12}^{K,Z'}\big)+\Delta M_K^{\rm LD}$\, and
\,$|\epsilon_K^{}|=\big|{\rm Im}\big(M_{12}^{K,\rm SM}+M_{12}^{K,Z'}\big)\big|/
\big(\sqrt2\,\Delta M_K^{\rm exp}\big)$,\,
where $\Delta M_K^{\rm LD}$ encodes long-distance effects and
\,$\Delta M_K^{\rm exp}=(52.89\pm0.10)\!\times\!10^{10}$/s\, \cite{pdg}.
Given the potentially sizable uncertainties in the $\Delta M_K$
calculation~\cite{Buras:2012jb}, we focus on $|\epsilon_K^{}|$, whose measured and SM values
are \,$|\epsilon_K^{\rm exp}|=(2.228\pm0.011)\times10^{-3}$ \cite{pdg} and
\,$|\epsilon_K^{\rm SM}|=\big(2.27_{-0.42}^{+0.21}\big)\times10^{-3}$~\cite{ckmfit}.
The 2$\sigma$ ranges of these numbers then suggest that we can impose
\,$\big|{\rm Im}\,M_{12}^{K,Z'}\big|<5\sqrt2\times10^{-4}\,\Delta M_K^{\rm exp}$,\,
which implies \,$\big|\zeta_1^{~}y_t^2+\zeta_2^{~}y_t^4\big|/m_{Z'}^{}<0.17$/TeV.

The flavor-changing $Z'$ couplings to $(d,s,b)$ affect the transition \,$b\to s\gamma$\,
via loop diagrams.
It is the best measured of \,$q\to q'\gamma$\, processes, with
\,${\cal B}(b\to s\gamma)_{\rm exp}^{}=(3.32\pm0.15) \times 10^{-4}$~\cite{hfag}
in agreement with the SM value
\,${\cal B}(b\to s\gamma)_{\textsc{sm}}^{}=(3.36\pm0.23)\times10^{-4}$ \cite{Misiak:2015xwa}.
Based upon these numbers, our computation of the $Z'$ effect on \,$b\to s\gamma$\,
leads to a constraint far weaker than~Eq.\,\eqref{zyt}, confirming earlier findings
in the literature~\cite{Buras:2012jb,Gauld:2013qja}.

\subsection{Four-Lepton Interactions \label{sec:4l}}

The ${\cal O}^{4\ell}$ operator in ${\cal L}_{\textsc{mfv}}$, induced by the $Z'$ boson at
tree level, gives rise to various processes that conserve or violate lepton flavor at tree
level or 1-loop level.
As searches for the flavor-violating decays of charged leptons have yielded the most
stringent bounds on some of the interactions of interest, we treat these processes first.

For \,$\ell_1\to\ell_2\ell_3\bar\ell_4$\, and \,$\ell_1\to\ell_2\gamma$,\, we employ
the relevant formulas from Ref.~\cite{Chiang:2011cv}.
Thus, we arrive at the rates
\begin{eqnarray}
\Gamma_{\tau\to ee\bar\mu}^{} &=& \frac{ \big| (\Delta_\ell)_{12}^{}
(\Delta_\ell)_{13}^{} \big|^2 m_\tau^5}{768_{\,}\pi^3\,m_{Z'}^4} \,, ~~~~ ~~~
\Gamma_{\tau\to\mu\mu\bar e}^{} \,=\, \frac{ \big| (\Delta_\ell)_{21}^{}
(\Delta_\ell)_{23}^{} \big|^2 m_\tau^5}{768_{\,}\pi^3\,m_{Z'}^4} \,,
\nonumber \\
\Gamma_{\tau\to\mu e\bar e}^{} &=& \frac{ \big| (\Delta_\ell)_{21}^{} (\Delta_\ell)_{13}^{}
\big|^2 m_\tau^5} {1536_{\,}\pi^3\,m_{Z'}^4} \,, ~~~~ ~~~
\Gamma_{\tau\to3\mu}^{} \,=\, \frac{ \big| (\Delta_\ell)_{22}^{}(\Delta_\ell)_{23}^{}
\big|^2 m_\tau^5}{768_{\,}\pi^3\,m_{Z'}^4} \,,
\nonumber \\
\Gamma_{\tau\to e\mu\bar\mu}^{} &=&
\frac{ \big|(\Delta_\ell)_{22}^{}(\Delta_\ell)_{13}^{} + (\Delta_\ell)_{12}^{}
(\Delta_\ell)_{23}^{} \big|^2 m_\tau^5} {1536_{\,}\pi^3\,m_{Z'}^4}
\end{eqnarray}
from tree-level $Z'$-exchange diagrams and
\begin{eqnarray}
\Gamma_{\mu\to e\gamma}^{} &=& \frac{\alpha_{\rm e\,}^{}m_\mu^5}
{2304_{\,}\pi^4\,m_{Z'}^4} \big| (\Delta_\ell)_{12}^{}(\Delta_\ell)_{22}^{} +
(\Delta_\ell)_{13}^{}(\Delta_\ell)_{32}^{} \big|^2 \,,
\nonumber \\
\Gamma_{\tau\to e\gamma}^{} &=& \frac{\alpha_{\rm e\,}^{}m_\tau^5}
{2304_{\,}\pi^4\,m_{Z'}^4} \big| (\Delta_\ell)_{12}^{}(\Delta_\ell)_{23}^{}
+ (\Delta_\ell)_{13}^{}(\Delta_\ell)_{33}^{} \big|^2 \,,
\nonumber \\
\Gamma_{\tau\to\mu\gamma}^{} &=& \frac{\alpha_{\rm e\,}^{}m_\tau^5}{2304_{\,}\pi^4\,m_{Z'}^4}
\big| (\Delta_\ell)_{21}^{}(\Delta_\ell)_{13}^{} + (\Delta_\ell)_{22}^{}(\Delta_\ell)_{23}^{}
+ (\Delta_\ell)_{23}^{}(\Delta_\ell)_{33}^{} \big|^2
\end{eqnarray}
from $Z'$-loop diagrams, where we have neglected the final leptons' masses and taken into
account the choice \,$(\Delta_\ell)_{11}=0$,\, which also leads to
\,$\Gamma_{\mu\to3e}=\Gamma_{\tau\to3e}=0$.\,
The experimental data are~\cite{pdg,meg}
\begin{eqnarray} \label{l->3l'}
{\cal B}(\tau\to ee\bar\mu)_{\rm exp}^{} &<& 1.5\times10^{-8} \,, ~~~~~
{\cal B}(\tau\to\mu\mu\bar e)_{\rm exp}^{} \,<\, 1.7\times10^{-8} \,,
\nonumber \\
{\cal B}(\tau\to e\mu\bar\mu)_{\rm exp}^{} &<& 2.7\times10^{-8} \,, ~~~~~
{\cal B}(\tau\to 3\mu)_{\rm exp}^{} \,<\, 2.1\times10^{-8} \,,
\nonumber \\
{\cal B}(\tau\to\mu e\bar e)_{\rm exp}^{} &<& 1.8\times10^{-8} \,, ~~~~~
{\cal B}(\mu\to e\gamma)_{\rm exp}^{} \,<\, 4.2\times10^{-13} \,,
\nonumber \\
{\cal B}(\tau\to e\gamma)_{\rm exp}^{} &<& 3.3\times10^{-8} \,, ~~~~~
{\cal B}(\tau\to\mu\gamma)_{\rm exp}^{} \,<\, 4.4\times10^{-8} \,,
\end{eqnarray}
all at 90\% CL.
The strictest of the bounds on these decay modes is from
${\cal B}(\mu\to e\gamma)_{\rm exp}^{}$, which translates into
\begin{eqnarray} \label{m2eg}
\frac{\big|(\Delta_\ell)_{12}^{}(\Delta_\ell)_{22}^{} +
(\Delta_\ell)_{13}^{}(\Delta_\ell)_{32}^{}\big|}{m_{Z'}^2}
\,<\, \frac{5.4\times10^{-4}}{\rm TeV\raisebox{0.3pt}{$^2$}} \,.
\end{eqnarray}
This indicates that some tuning is needed so that
\,$(\Delta_\ell)_{22}^{}/m_{Z'}^{}=\cal O$(0.2)/TeV\,
can be maintained in order to satisfy Eq.~(\ref{ceecmm}).
The other modes, notably \,$\tau\to3\mu$,\, can also be important.

Related to \,$\ell_1\to\ell_2\gamma$\, is the $Z'$ contribution to the anomalous magnetic
moment of charged lepton $\ell_j$,
\begin{eqnarray} \label{alZ'}
a_{\ell_j}^{Z'} \,\simeq\, \frac{-m_{\ell_j}^2}{12\pi^2\,m_{Z'}^2}\,
\raisebox{2pt}{\footnotesize$\displaystyle\sum_k$}\,
\big|(\Delta_\ell)_{jk}^{}\big|\raisebox{2pt}{$^2$} \,.
\end{eqnarray}
With $a_{\ell_j}^{Z'}$ being always negative, due to the $Z'$ in this study possessing
purely left-handed fermionic couplings, it does not help resolve the discrepancy
between $a_\mu^{\textsc{sm}}$ and
$a_\mu^{\rm exp}$, presently differing by
\,$a_\mu^{\rm exp}-a_\mu^{\textsc{sm}}=(288\pm80)\times10^{-11}$~\cite{pdg}.
Thus, if confirmed in the future to have a NP origin, the deviation would need to be
explained with extra ingredients beyond our specific $Z'$ scenario.
Nevertheless, requiring $\big|a_{\ell_j}^{Z'}\big|$ to be less than the error in
this difference does not result in a strict limitation on the $Z'$ couplings.

The $Z'$-loop diagrams responsible for $a_{\ell_j}^{Z'}$ generally also impact the electric
dipole moment (EDM) of $\ell_j$.
However, with the pertinent formula from Ref.~\cite{Chiang:2011cv}, it is straightforward
to realize that, the $Z'$ having purely left-handed fermionic couplings, its contribution to
the EDM of $\ell_j$ vanishes exactly at the 1-loop level.
For the same reason, our $Z'$ has no effect on the EDM of a~quark.

Another type of low-energy process which can be affected by the $Z'$ is the SM-dominated
decay \,$\ell\to\ell'\nu\nu'$.\,
Since the neutrinos are unobserved, its rate comes from channels with all
possible combinations of neutrino flavors in the final states, namely
\begin{eqnarray} \label{Gt2enn'}
\Gamma_{\tau\to\mu\nu\nu'} \,=\, \Gamma_{\tau\to\mu\nu_\mu^{}\nu_\tau^{}} +
\Gamma_{\tau\to\mu\nu_e^{}\nu_\mu^{}} + \Gamma_{\tau\to\mu\nu_e^{}\nu_\tau^{}} +
\Gamma_{\tau\to\mu\nu_\tau^{}\nu_\tau^{}} + \Gamma_{\tau\to\mu\nu_\mu^{}\nu_\mu^{}} \,,
\end{eqnarray}
where
\begin{eqnarray}
\Gamma_{\tau\to\mu\nu_\mu^{}\nu_\tau^{}} &=& \frac{G_{\rm F}^2\,m_\tau^5}{192\pi^3}
\big(1-8\rho_\mu^{}+8\rho_\mu^3-\rho_\mu^4-12\rho_\mu^2\,\ln\rho_\mu^{}\big) (1+R_{23})^2 \,,
\nonumber \\
\rho_\ell^{} &=& \frac{m_\ell^2}{m_\tau^2} \,, ~~~~~
R_{rs}^{} \,=\, \frac{|(\Delta_\ell)_{rs}^{}|^2}{\sqrt8~G_{\rm F\,}^{}m_{Z'}^2} \,,
\end{eqnarray}
and the other partial rates in Eq.~(\ref{Gt2enn'}) can be neglected, being without SM
contributions and proportional to \,$\big|(\Delta_\ell)_{23}^2(\Delta_\ell)_{rs}^2\big|$.\,
One could write down an analogous formula for $\Gamma_{\tau\to e\nu\nu'}$.
From the data \,${\cal B}(\tau\to e\nu\nu')_{\rm exp}^{}=(17.82\pm0.04)\%$\, and
\,${\cal B}(\tau\to\mu\nu\nu')_{\rm exp}^{}=(17.39\pm0.04)\%$ \cite{pdg}
and SM predictions
\,${\cal B}(\tau\to e\nu\nu')_{\textsc{sm}}^{}=0.1778\pm0.0003$\, and
\,${\cal B}(\tau\to\mu\nu\nu')_{\textsc{sm}}^{}=0.1729\pm0.0003$ \cite{Pich:2013lsa},
we calculate
\begin{eqnarray} \label{tau2lnn}
\frac{{\cal B}(\tau\to e\nu\nu')_{\rm exp}^{}}{{\cal B}(\tau\to e\nu\nu')_{\textsc{sm}}^{}}
\,=\, 1.002\pm0.006  \,, ~~~~~~~
\frac{{\cal B}(\tau\to\mu\nu\nu')_{\rm exp}^{}}{{\cal B}(\tau\to\mu\nu\nu')_{\textsc{sm}}^{}}
\,=\, 1.006\pm0.006 \,,
\end{eqnarray}
with 2$\sigma$ errors.
Numerically, we get \,$(1+R_{13,23})^2-1<0.0011$\, for the $Z'$ effect represented by our
benchmark points, and so it is at least several times smaller than the errors in
Eq.\,(\ref{tau2lnn}).

At higher energies, the $Z'$ contributions may be probed by LEP experiments on the scattering
\,$e^+e^-\to\ell^+\ell^-$\, for \,$\ell=e,\mu,\tau$.\,
In particular, the lower limits at 95\% CL on the effective heavy mass scale
derived from fits to their data~\cite{Alcaraz:2006mx} imply
\begin{eqnarray}
\frac{\big|(\Delta_\ell)_{11}^{}\big|}{m_{Z'}^{}} \,\le\, \frac{0.28}{\rm TeV} \,, ~~~~ ~~~
0 \,\le\, \frac{(\Delta_\ell)_{11}^{}(\Delta_\ell)_{jj}^{}
+ (\Delta_\ell)_{1j}^{}(\Delta_\ell)_{j1}^{}}{m_{Z'}^2} \,\le\,
\frac{0.13}{\rm TeV\raisebox{0.3pt}{$^2$}} \,,
\end{eqnarray}
where \,$j=2,3$.\,
The first constraint is automatically satisfied by our preference
\,$(\Delta_\ell)_{11}^{}=0$,\, and consequently, since $\Delta_\ell$ is Hermitian,
the second one becomes
\begin{eqnarray}
\frac{\big|(\Delta_\ell)_{1j}^{}\big|}{m_{Z'}^{}} \,\le\, \frac{0.36}{\rm TeV} \,, ~~~~~
j \,=\, 2,3 \,.
\end{eqnarray}
As can be expected, these restrictions turn out to be less important than that in Eq.\,(\ref{m2eg}).
Although not explicitly addressed in this study, we mention that the leptonic $Z'$ couplings
contribute at 1-loop level to $Z$-pole observables, such as the $Z$ leptonic partial-decay rates
and forward-backward asymmetries, also measured at LEP~\cite{pdg},
but the implied restraints are not strong either, provided that \,$m_{Z'}^{}>0.5$\,TeV.\,

\section{Numerical Analysis \label{sec:analysis}}

As discussed in the preceding sections, we deal with the fermionic interactions of the $Z'$
by imposing MFV on both its quark and lepton couplings and, for the latter, by
incorporating the type-I seesaw mechanism with 3 heavy right-handed neutrinos.
One could perform instead a\,\,simpler implementation of leptonic MFV by assuming
a minimal field content with only SM fermions plus the dimension-5 Weinberg operator,
as was done in Ref.\,\,\cite{Kim:2016bdu}.
However, in the type-I seesaw case, there is significantly more freedom to satisfy the various
constraints.

Given that the $Z'$ leptonic interactions in Eq.\,(\ref{LZ'}) involve
\,$\Delta_\ell=\xi_0^{}\openone+\xi_1^{~}\textsf{A}_\ell^{}+\xi_{2\,}^{}\textsf{A}_\ell^2$\,
with $\textsf{A}_\ell$ defined in Eq.\,(\ref{AqAl}), to evaluate them we need the values of
the elements of $U_{\textsc{pmns}}$, $\hat m_\nu$, and $OO^\dagger$,
as well as the coefficients $\xi_{0,1,2}^{}$.
Thus, for $U_{\rm PMNS}$, adopting the standard parametrization~\cite{pdg}, we employ
the parameter values quoted in Table\,\ref{nu-data} from a recent fit to global neutrino
data~\cite{nudata}.
The majority of these numbers depend on whether the light neutrinos' masses have a normal
ordering (NO), where \,$m_1<m_2<m_3$,\, or an inverted one (IO), where \,$m_3<m_1<m_2$.\,
As the absolute scale of $m_{1,2,3}$ is not yet established, for definiteness we will pick
\,$m_{1(3)}=0$\, in the NO (IO) case.
In general $U_{\textsc{pmns}}$ may also contain Majorana phases, which are still
unknown, but for simplicity we set them to zero.
As for $\xi_{0,1,2}^{}$, one of them is no longer free due to the requisite
\,$(\Delta_\ell)_{11}^{}=0$\, implied by Eq.\,(\ref{ceecmm}).
This allows us to fix
\,$\xi_0^{}=-\xi_1^{~}(\textsf{A}_\ell)_{11}^{}-\xi_2^{~}\big(\textsf{A}_\ell^2\big)_{11}$,\,
but permit the other two coefficients to have any real values as long as
\,$\big|\xi_{1,2}^{}\big|\le\cal O$(1).\,

\begin{table}[b] \medskip
\begin{tabular}{|c||c|c|} \hline Parameter & NO & IO$\vphantom{|^{|^|}}$\\ \hline\hline
${\rm sin}^2\theta_{12}^{}$ & $0.306\pm0.012\vphantom{\frac{1}{2}_|}$ & $0.306\pm0.012$
\\
${\rm sin}^2\theta_{23}^{}$ & $0.441_{-0.021}^{+0.027}$ &
$0.587_{-0.024}^{+0.020}\vphantom{\frac{1}{2}_|}$
\\
${\rm sin}^2\theta_{13}^{}$ & $0.02166\pm0.00075$ & $0.02179\pm0.00076\vphantom{\frac{1}{2}_|}$
\\
$\delta/{}^\circ$ & $261_{-59}^{+51}$ & $277_{-46}^{+40}\vphantom{\frac{1}{2}_|}$
\\
\,$\Delta m_{21}^2 = m_2^2-m_1^2$\, &
$\left(7.50_{-0.17}^{+0.19}\right)_{\vphantom{\int}}^{\vphantom{\int}}\!\times\!10^{-5}\;\rm eV^2$
& $\left(7.50_{-0.17}^{+0.19}\right)\times10^{-5}\;\rm eV^2$
\\
~$\Delta m_{3\ell}^2\vphantom{\int_{|_|}^{|}}$~ &
~$m_3^2-m_1^2=\big(2.524_{-0.040}^{+0.039}\big)\!\times\! 10^{-3}\;\rm eV^2$~ &
~$m_3^2-m_2^2=\bigl(-2.514_{-0.041}^{+0.038}\big)\!\times\! 10^{-3}\;\rm eV^2$~
\\ \hline
\end{tabular} \smallskip
\caption{The best-fit values, and their one-sigma errors, of neutrino oscillation parameters
from the global analysis in Ref.\,\,\cite{nudata}.
The entries under NO (IO) correspond to the normal (inverted) ordering of the light neutrinos'
masses.\label{nu-data}}
\end{table}

In our numerical explorations, we vary the neutrino quantities listed in Table\,\,\ref{nu-data}
within their 2$\sigma$ intervals and confine $\xi_{1,2}^{}/m_{Z'}^{}$ to between $\pm1.5/$TeV.
To help ensure perturbativity, we always require the biggest eigenvalue of $\textsf{A}_\ell$
equal unity, which implies that the right-handed neutrinos' mass $\cal M$ is of order
$10^{13}$-$10^{15}$\,\,GeV in our examples.
Furthermore, to optimize the size of $\textsf{c}_{\mu\mu}$ according to Eq.\,(\ref{cll'}),
we select \,$\big(\zeta_1^{~}y_t^2+\zeta_2^{~}y_t^4\big)/m_{Z'}^{}=0.13$/TeV,\,
which is the maximum as dictated by Eq.\,(\ref{zyt}).

We begin our numerical analysis by looking first at the simplest possibility for $\textsf{A}_\ell$
in Eq.\,(\ref{AqAl}), which is that the orthogonal matrix $O$ is real and hence
\,$\textsf{A}_\ell=2{\mathcal M}U_{\textsc{pmns}\,}^{}\hat m_\nu
U_{\textsc{pmns}}^\dagger/v^2$.\,
Upon scanning the parameter space in this scenario subject to the restrictions detailed above,
for the NO case we find that we can attain
\,$-0.46\;\mbox{\footnotesize$\lesssim$}\;\textsf{c}_{\mu\mu}\le-0.32$,\, which is
a portion of the $\textsf{c}_{\mu\mu}$ range in Eq.\,(\ref{ceecmm}), but on its upper side,
as long as the Dirac $CP$-violation phase $\delta$ in $U_{\textsc{pmns}}$ lies below
its central value in Table\,\,\ref{nu-data} by about 1$\sigma$ or more.
Consequently, although it may be too early to rule out this possibility, it is disfavored.
The status of the IO case is worse, as we are not able to reach the desired values of
$\textsf{c}_{\mu\mu}$ during our scans.
The limitations on these cases are caused partly by the small value of
\,$\big(\zeta_1^{~}y_t^2+\zeta_2^{~}y_t^4\big)/m_{Z'}^{}$\, picked in the last paragraph.
Another reason is that $(\Delta_\ell)_{22}$ is also small because it has only two free parameters,
$\xi_{1,2}^{}$, which are subject mainly to the strict empirical bounds on charged-LFV decays,
especially \,$\mu\to e\gamma$.\,
It is therefore of interest to consider another choice of ${\textsf A}_\ell$ which has
a less simple structure, but which may offer additional adjustable parameters.

A more promising situation is when $\textsf{A}_\ell$ in Eq.\,\,\eqref{AqAl} contains a complex $O$ matrix.
Since we can in general write \,$O=e^{i\sf R}e^{{\sf R}'}$ with real antisymmetric
matrices $\sf R$ and ${\sf R}'$, we have
\begin{eqnarray} \label{oo+}
OO^\dagger \,=\, e^{2i\sf R} \,, ~~~~~~~
{\sf R} \,= \left(\!\begin{array}{ccc} 0 & r_1^{} & r_2^{} \vspace{1pt} \\
-r_1^{} & 0 & r_3^{} \vspace{1pt} \\ -r_2^{} & \mbox{$-r_3^{}$} & 0 \end{array}\!\right) ,
\end{eqnarray}
where $r_{1,2,3}$ are independent real constants.
These extra free parameters prove to be advantageous for our purposes.
When conducting our scans in this scenario, we let the other parameters fall within their
ranges specified before in this section, whereas $r_{1,2,3}$ are allowed to have any real
values.\footnote{In our numerical analysis, we aim mainly at obtaining viable solutions
under our MFV framework with the $Z'$ that can account for the \,$b\to s\ell^+\ell^-$\,
anomalies and looking at some of the implications.
As our results demonstrate, there are indeed a substantial amount of points in the $Z'$
parameter space of interest which can accomplish our purposes and are simultaneously
compatible with the pertinent constraints.
Therefore, in this study, as also in \cite{Lee:2015qra}, we leave aside concerns about
the issue of fine tuning which has been raised in \cite{Dinh:2017smk}.}

\begin{figure}[t]
\includegraphics[width=4in]{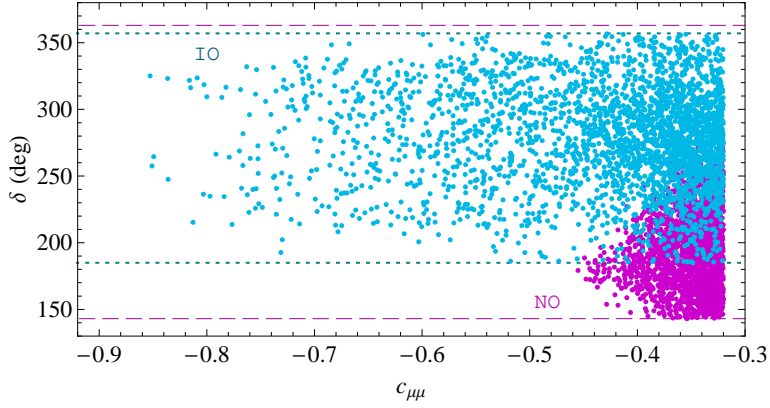}\vspace{-7pt}
\caption{Distributions of the Dirac $CP$-violation phase $\delta$ in $U_{\textsc{pmns}}$
versus $\textsf{c}_{\mu\mu}$ corresponding to benchmark points within the allowed parameter
space in the NO (magenta) and IO (cyan) cases.
The magenta dashed (cyan dotted) lines mark the boundaries of the 2$\sigma$ region of
$\delta$ in the NO (IO) case.} \label{delta-cmm-plots}
\end{figure}

With $O$ being complex, during our scans we can obtain ${\textsf c}_{\mu\mu}$ values consistent
with Eq.\,(\ref{ceecmm}) and at the same time all the neutrino mixing parameters can stay within
their 2$\sigma$ regions, including $\delta$ which can fall even inside its 1$\sigma$ range.
To illustrate this, in Fig.\,\ref{delta-cmm-plots} we present sample distributions of $\delta$
versus $\textsf{c}_{\mu\mu}$ in the NO (magenta) and IO (cyan) cases corresponding, respectively,
to 2000 and 3000 benchmark points in the parameter space fulfilling the different constraints
described earlier.
Evidently, it is easier in the IO scenario to achieve a larger size of $\textsf{c}_{\mu\mu}$
while satisfying the various restrictions.
This appears to be the opposite of what we saw in the real-$O$ case and may simply have to do
with the current neutrino and other lepton data situation which could still change in the future.

\begin{figure}[!t] \bigskip
\includegraphics[width=53mm]{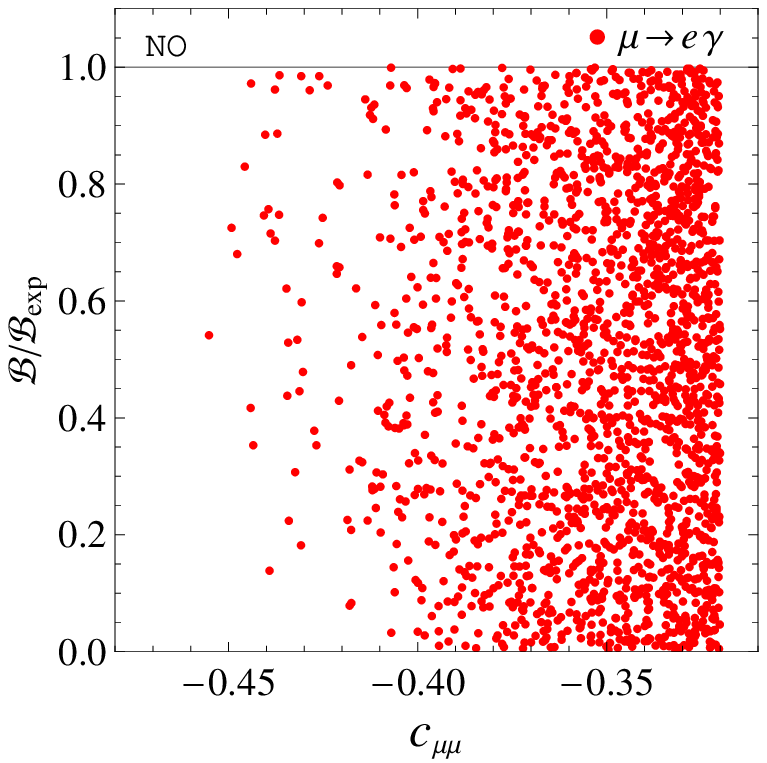}
\includegraphics[width=53mm]{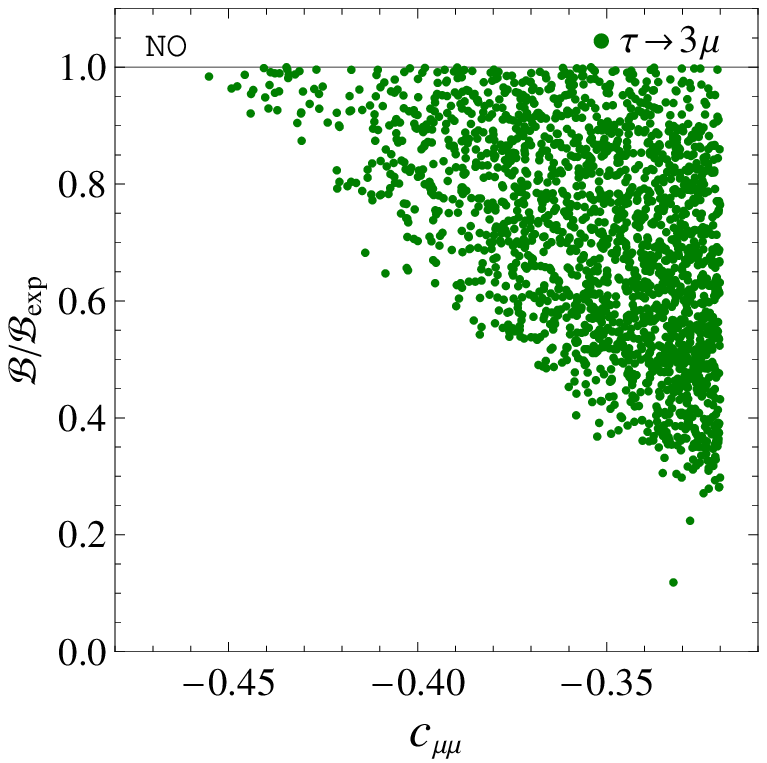}
\includegraphics[width=53mm]{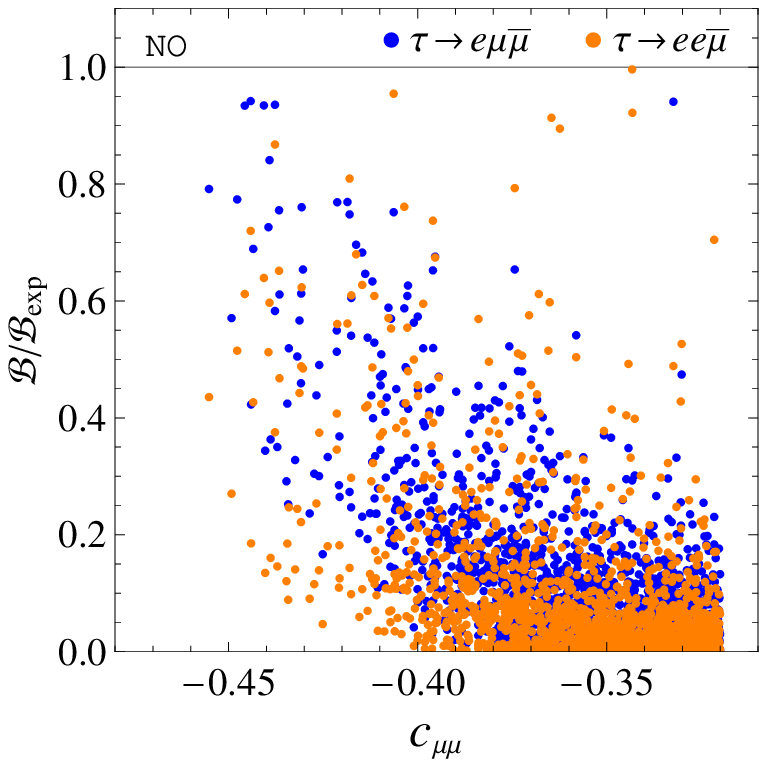}\vspace{-7pt}
\caption{Distributions of the branching fractions of \,$\mu\to e\gamma$\, and
\,$\tau\to3\mu,ee\bar\mu,e\mu\bar\mu$,\, divided by their respective experimental upper-limits,
versus $\textsf{c}_{\mu\mu}$ corresponding to the aforementioned benchmark points in the NO
case.} \label{no-plots}
\end{figure}

\begin{figure}[!t] \bigskip
\includegraphics[width=53mm]{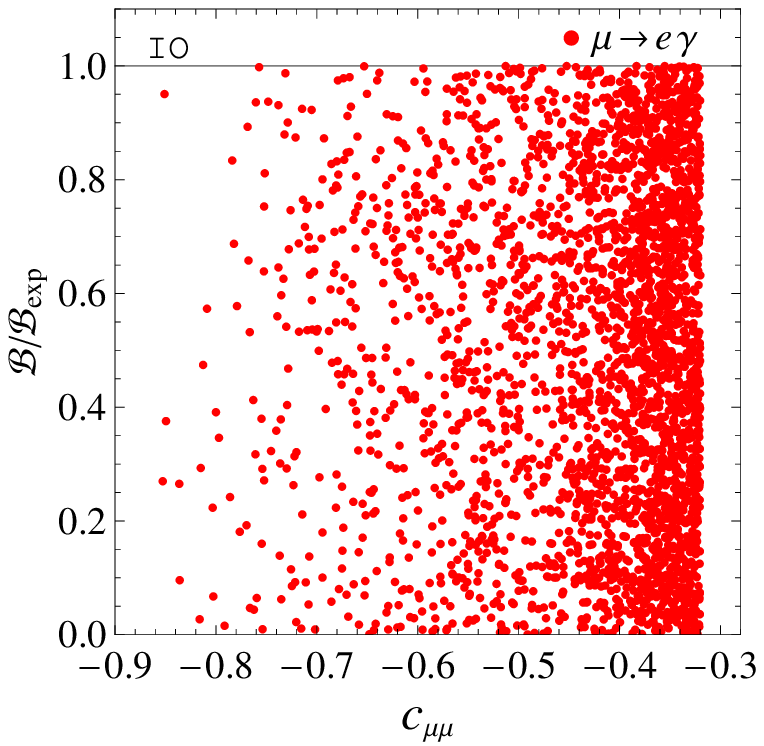}
\includegraphics[width=53mm]{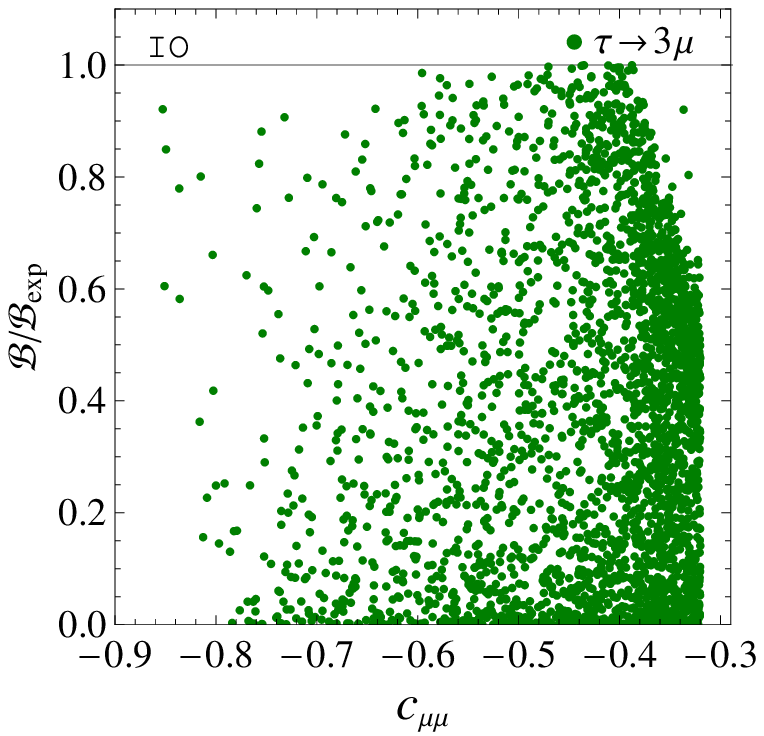}
\includegraphics[width=53mm]{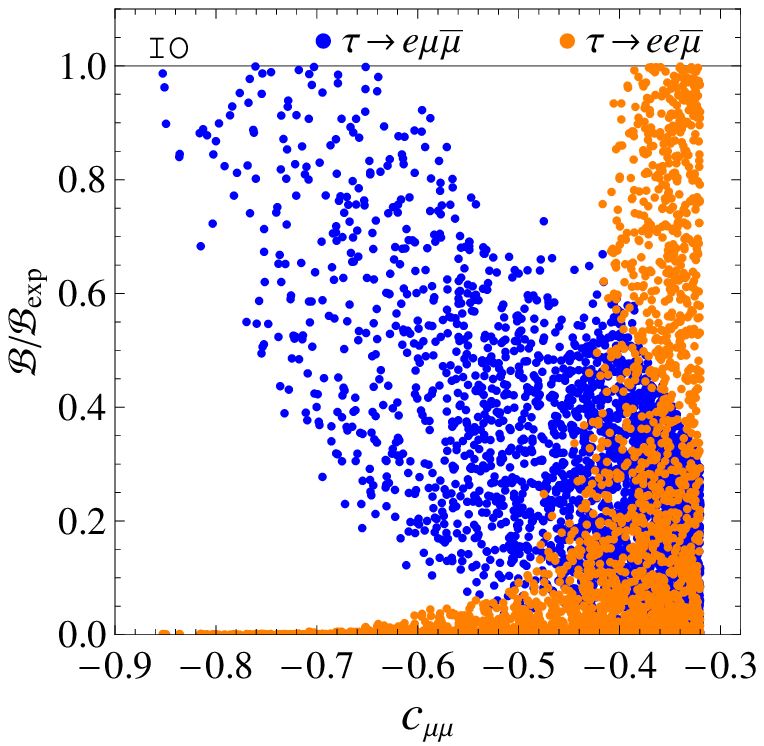}\vspace{-7pt}
\caption{The same as Fig.\,\ref{no-plots}, but for the IO case.} \label{io-plots}
\end{figure}

As expected, the limit from \,$\mu\to e\gamma$\, searches plays a major constraining role
for many of the benchmarks, as can be viewed in Figs.\,\,\ref{no-plots} and\,\,\ref{io-plots},
where we plot the branching fractions of
\,$\mu\to e\gamma$\, and \,$\tau\to3\mu,ee\bar\mu,e\mu\bar\mu$\, normalized by their
respective experimental bounds, which are quoted in\,\,Eq.\,(\ref{l->3l'}),
versus $\textsf{c}_{\mu\mu}$.
The \,$\tau\to 3\mu$\, data can also be important, especially in the NO case, in which
\,$\textsf{c}_{\mu\mu}<-0.46$\, is not possible without ${\cal B}(\tau\to3\mu)$ violating
its empirical limit, as can be inferred from the middle plot in Fig.\,\ref{no-plots}.
In these figures, we do not display the corresponding ratios for
\,$\tau\to e\gamma,\mu\gamma,\mu\mu\bar e,\mu e\bar e$\, because they are
comparatively less able to reach unity.

\section{Predictions \label{sec:predictions}}

We notice in Figs.\,\,\ref{no-plots} and\,\,\ref{io-plots} that there exists parameter space
where the branching fractions of the various LFV decays approach their current experimental limits,
even within factors of a few, while a sizeable $c_{\mu\mu}$ is still allowed.
They are testable with future quests or detections of these charged-LFV decays and with upcoming
improved measurements of \,$b\to s\ell^+\ell^-$\, processes.

\begin{figure}[b]
\includegraphics[width=4in]{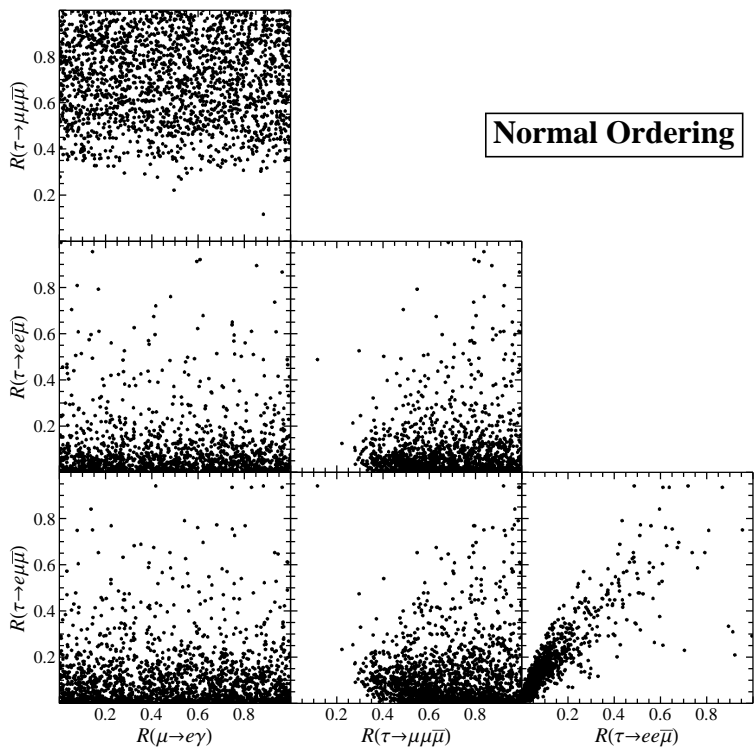}\vspace{-7pt}
\caption{Distributions of pairs of the ratios \,$R={\cal B}/{\cal B}_{\rm exp}$\, shown in
Fig.\,\ref{no-plots} for the different LFV decay channels in the NO case.} \label{correlations-no}
\end{figure}

\begin{figure}[t]
\includegraphics[width=4in]{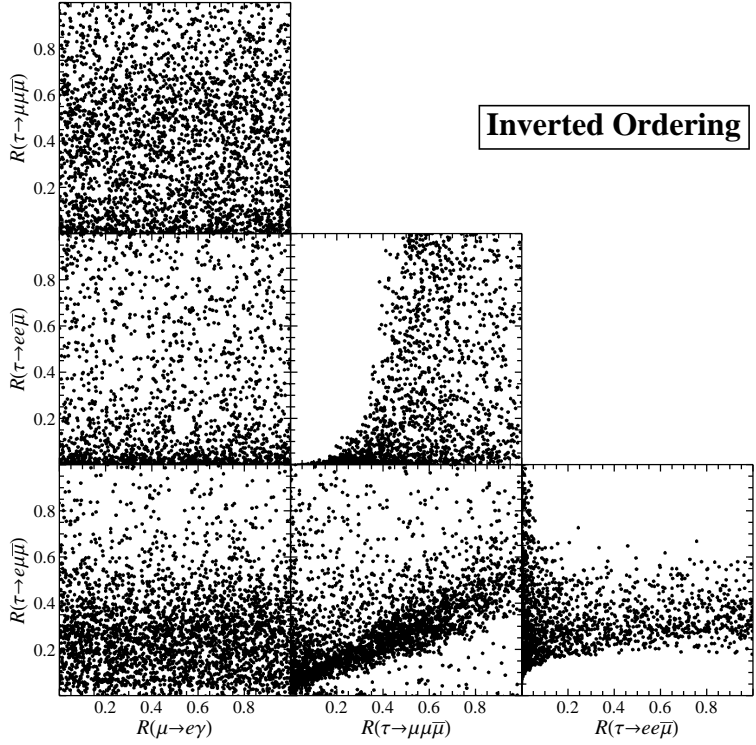}\vspace{-7pt}
\caption{Distributions of pairs of the ratios \,$R={\cal B}/{\cal B}_{\rm exp}$\, shown in
Fig.\,\ref{io-plots} for the different LFV decay channels in the IO case.} \label{correlations-io}
\end{figure}

In Figs.\,\,\ref{no-plots} and\,\,\ref{io-plots}, we also see that the NO and IO scenarios
predict different potential correlations among the branching fractions of these decays which
may be confirmed or excluded when they are observed in the future with sufficient precision.
To illustrate these possibilities, based on those graphs we present in
Figs.\,\,\ref{correlations-no} and\,\,\ref{correlations-io} the distributions of
several pairs of the ratios \,$R={\cal B}/{\cal B}_{\rm exp}$\, of the calculated branching
fractions to their respective experimental bounds.

The fact that Eq.\,(\ref{cll'}) also describes LFV couplings implies that they give rise to
\,$b\to s\ell\bar\ell'$\, and, analogously, also \,$b\to d\ell\bar\ell'$\, and
\,$s\to d\ell\bar\ell'$,\, with \,$\ell\neq\ell'$,\, all of which strictly do not occur in
the SM with massless neutrinos.
Accordingly, we have predictions for a number of exclusive $B_{d,s}$-meson and kaon decays.
Using the pertinent formulas given in Ref.\,\cite{Lee:2015qra}, with updated CKM
parameters~\cite{ckmfit}, we determine the maximum $|\textsf{c}_{\ell\ell'}|$ from our benchmark
points to calculate the branching fractions collected in Table\,\,\ref{Cetmt}.
We observe that the predictions for a few of the modes ({\it e.g.}, \,$B\to K^{(*)}e\mu$,
\,$B\to\pi e\mu$,\, and \,$K_L\to e\mu$) are within two orders of magnitude from
their experimental bounds, especially \,$K_L\to e\mu$,\, and consequently may be
probed in near-future searches.

\begin{table}[t]
\begin{tabular}{|c||c|c|c|} \hline
\multirow{3}{*}{Decay mode} & \multicolumn{3}{c|}{Branching fractions} \\ \cline{2-4} &
~Measured upper limit~ & \multicolumn{2}{c|}{Prediction maximum [or range]} \\ \cline{3-4} &
at 90\% CL~\mbox{\cite{pdg,hfag}} & NO & IO \\ \hline\hline
$B\to K e^\pm\mu^\mp\vphantom{|^{\int^|}}$  & $3.8\times10^{-8}$ & $2.9\times10^{-9}$ &
$3.0\times10^{-9}$
\\
$B\to K^*e^\pm\mu^\mp\vphantom{|^{\int^|}}$ & $5.1\times10^{-7}$ & $7.8\times10^{-9}$ &
$7.8\times10^{-9}$
\\
$B_s\to e^\pm\mu^\mp\vphantom{|^{\int^|}}$  & $1.1\times10^{-8}$ & $8.6\times10^{-12}$ &
~ $9.0\times10^{-12}\vphantom{|_{|_O^{}}^{}}$ ~
\\ \hline
$B\to\pi e^\pm\mu^\mp\vphantom{|^{\int^|}}$ & $9.2\times10^{-8}$ & $1.2\times10^{-10}$ &
$1.3\times10^{-10}$
\\
$B\to\rho e^\pm\mu^\mp\vphantom{|^{\int^|}}$ & $3.2\times10^{-6}$  & $3.1\times10^{-10}$ &
$3.2\times10^{-10}$
\\
$B^0\to e^\pm\mu^\mp\vphantom{|^{\int^|}}$   & $2.8\times10^{-9}$  & $2.6\times10^{-13}$ &
$2.7\times10^{-13}\vphantom{|_{|_O^{}}^{}}$
\\ \hline
$B^+\to K^+e^\pm\tau^\mp\vphantom{|^{\int^|}}$    & $3.0\times10^{-5}$ & $8.1\times10^{-9}$ &
$5.9\times10^{-9}\vphantom{|_{|_O^{}}^{}}$
\\
~$B^+\to K^{*+}e^\pm\tau^\mp\vphantom{|^{\int^|}}$~ & -- & $1.6\times10^{-8}$    &
$1.2\times10^{-8}$
\\
$B_s\to e^\pm\tau^\mp\vphantom{|^{\int^|}}$   & -- & $8.0\times10^{-9}$    &
$5.8\times10^{-9}\vphantom{|_{|_O^{}}^{}}$
\\ \hline
$B^+\to\pi^+e^-\tau^+\vphantom{|^{\int^|}}$   & $2.0\times10^{-5}$ & $1.9\times10^{-10}$ &
$1.4\times10^{-10}$
\\
$B^+\to\rho^+e^\pm\tau^\mp\vphantom{|^{\int^|}}$  & -- & $7.1\times\!10^{-10}$   &
$5.2\times\!10^{-10}$
\\
$B^0\to e^\pm\tau^\mp\vphantom{|^{\int^|}}$ & $2.8\times10^{-5}$ & ~$2.4\times\!10^{-10}$~ &
$1.7\times10^{-10}\vphantom{|_{|_O^{}}^{}}$
\\ \hline
$B^+\to K^+\mu^\pm\tau^\mp\vphantom{|^{\int^|}}$ & $4.8\times10^{-5}$ &
$[0.3,3.1]\!\times\!10^{-9}$ & $2.6\times10^{-9}$
\\
$B^+\to K^{*+}\mu^\pm\tau^\mp\vphantom{|^{\int^|}}$  & $4.8\times10^{-5}$ &
$[0.7,6.1]\!\times\!10^{-9}$ & $5.1\times10^{-9}$
\\
$B_s\to\mu^\pm\tau^\mp\vphantom{|^{\int^|}}$ & -- & $[0.3,3.1]\!\times\!10^{-9}$ &
$2.6\times10^{-9}$
\\ \hline
$B^+\to\pi^+\mu^\pm\tau^\mp\vphantom{|^{\int^|}}$ & $7.2\times10^{-5}$ &
$[0.2,1.5]\!\times\!10^{-10}$    & $1.2\times10^{-10}$
\\
$B^+\to\rho^+\mu^\pm\tau^\mp\vphantom{|^{\int^|}}$ & $7.2\times10^{-5}$ &
~$[0.3,2.7]\!\times\!10^{-10}$~    &    $2.3\times10^{-10}$
\\
$B^0\to\mu^\pm\tau^\mp\vphantom{|^{\int^|}}$  & $2.2\times10^{-5}$ &
$[1,9]\!\times\!10^{-11}$    &    $7.7\times10^{-11}\vphantom{|_{|_O}^{}}$
\\ \hline
$K_L^{}\to e^\pm\mu^\mp\vphantom{|^{\int^|}}$ & $4.7\times10^{-12}$ & ~$1.4\times10^{-12}$~ &
$1.5\times10^{-12}\vphantom{|_{|_O}^{}}$
\\ \hline
\end{tabular} \smallskip
\caption{The maximum predictions for the branching fractions of exclusive $b$-meson (kaon) decays
involving  $e\mu$, $e\tau$, and $\mu\tau$ ($e\mu$) in the final states.
The lower end of a prediction is also displayed if exceeding one per mill of its upper end.
For comparison, the data are quoted if available.
To conform to the experimental reports~\cite{Edwards:2002kq}, the \,$B\to K^{(*)}e\mu$\,
prediction is the simple average over the $B^+$ and $B^0$ channels,
\,${\cal B}\big(B\to K^{{\scriptscriptstyle(}*\scriptscriptstyle)}e^\pm\mu^\mp\big)=
\big({\cal B}(B^+\to K^{{\scriptscriptstyle(}*{\scriptscriptstyle)}+}e^\pm\mu^\mp)
+ {\cal B}(B^0\to K^{{\scriptscriptstyle(}*{\scriptscriptstyle)}0}e^\pm\mu^\mp)\big)/2$,\,
whereas the \,$B\to\pi e\mu$\, prediction is from
\,${\cal B}\big(B\to\pi e^\pm\mu^\mp\big)={\cal B}(B^+\to\pi^+e^\pm\mu^\mp)/2
+ {\cal B}(B^0\to\pi^0e^\pm\mu^\mp)$\,
and similarly for \,$B\to\rho e^\pm\mu^\mp$.
The predictions for \,$B_s\to\phi\ell\ell'$\, are close to those for
\,$B\to K^*\ell\ell'$.} \label{Cetmt}
\end{table}

Future measurements of \,$b\to s\tau^+\tau^-$\, transitions, such as
\,$B\to K^{(*)}\tau^+\tau^-$,\, \,$B_s\to\phi\tau^+\tau^-$,\, and \,$B_{d,s}\to\tau^+\tau^-$,\,
all of which are not yet seen~\cite{hfag}, may be sensitive to the coefficient
${\textsf c}_{\tau\tau}$.
From our benchmarks, we derive \,$-0.63\,(-0.85)<\textsf{c}_{\tau\tau}^{}<+0.80\,(-0.11)$\,
in the NO (IO) case.
This implies that our $Z'$ scenario predicts a modification to the SM expectations of their
rates by a factor of
\begin{eqnarray} \label{b2stt}
0.72\,(0.64) \,<\, \Bigg(1+\frac{{\textsf c}_{\tau\tau}^{}}{\big|C_{9,10}^{\rm SM}\big|}
\Bigg)^{\!\!2}
\,<\, 1.42\,(0.95) \,.
\end{eqnarray}
Evidently the $Z^\prime$ impact on these decays can be fairly substantial, but experimental searches
for them are challenging due to elusive neutrinos being among the $\tau^\pm$ decay daughters.
For instance, the LHCb upgrade plan to collect a total data set of $50~{\rm fb}^{-1}$ can
improve upon the current bound \,${\cal B}(B_s\to\tau^+\tau^-)_{\rm exp}<5.2\times 10^{-3}$
at 90\% CL\,\,\cite{hfag,Aaij:2017xqt} to merely \,$5\times 10^{-4}$~\cite{Albrecht:2017odf},
which is far above the SM estimate of \,$7.6\times 10^{-7}$\,\,\cite{tb2stt}.
Similarly, \,${\cal B}(B^+\to K^+\tau^+\tau^-)_{\rm exp}<2.25\times10^{-3}$
at 90\% CL\,\,\cite{TheBaBar:2016xwe} may be improved upon in the Belle II experiment by no more
than two orders of magnitude~\cite{Kamenik:2017ghi}.
A much better situation could occur at a future $e^+e^-$ circular collider, the FCC-$ee$,
operating at the $Z$ pole, where full reconstructions of a few thousand
\,$B_d\to K^{*0}\tau^+\tau^-$\, events from ${\cal O}(10^{13})$ $Z$ decays would be potentially
achievable~\cite{Kamenik:2017ghi}, which might offer opportunities to probe the predictions
in Eq.\,\eqref{b2stt}.

Since the leptonic part of the operator ${\cal O}^{2q2\ell}$ in Eq.\,(\ref{Lmfv}) contains
light neutrinos besides the charged leptons, it contributes along with the SM to
\,$b\to(d,s)\nu\nu'$\, and \,$s\to d\nu\nu'$\, transitions.
Thus, their amplitudes involve the coefficients in Eq.\,(\ref{cll'}) as well.
Among the affected exclusive modes are \,$B\to(\pi,\rho)\nu\nu$,\, \,$B\to K^{(*)}\nu\nu$,\,
$K_L\to\pi^0\nu\nu$,\, and \,$K^+\to\pi^+\nu\nu$,\, all of which are subject to ongoing
experimental efforts~\cite{Ahn:2016kja,Lollini:2017kuw,Grygier:2017tzo} and only the last one
of which has been discovered, but with a significant uncertainty~\cite{pdg}.
Employing again the relevant formulas listed in Ref.\,\cite{Lee:2015qra}, from our benchmarks
points we estimate that in the NO (IO) case the rates of the $B$ and $K_L$ channels get altered
by a factor of \,$0.96\,(1.05)<r_{B\to(\pi,\rho,K,K^*)\nu\nu,K_L\to\pi^0\nu\nu}<1.11\,(1.19)$\,
and the $K^+$ channel by \,$0.97\,(1.03)<r_{K^+\to\pi^+\nu\nu}<1.08\,(1.13)$.
In the future, the KOTO~\cite{Komatsubara:2012pn} and NA62~\cite{Rinella:2014wfa} experiments are
expected to measure the rates of \,$K_L\to\pi^0\nu\nu$ and $K^+\to\pi^+\nu\nu$,\, respectively,
with about 10\% precision, and the proposed Project X experiment~\cite{Kronfeld:2013uoa} aims at
sensitivity of 5\% or less for their rates~\cite{Butler:2013kdw}.
Since the uncertainties of their SM rate predictions are currently around\,\,10\%, to detect
the above $\mathcal O(10\%)$ $Z^\prime$ effects on \,$K\to\pi\nu\nu$\, will require further
progress in theoretical efforts, such as improved lattice QCD calculations, and more precise
values of the CKM parameters~\cite{Buras:2015qea}.

\begin{table}[b] \medskip
\begin{tabular}{|c||c|c|c|c|c|} \hline
\multirow{4}{*}{$\begin{array}{c}\rm Decay \vspace{1pt} \\ \rm mode \end{array}$} &
\multicolumn{5}{c|}{Branching fractions} \\ \cline{2-6} &
\multirow{3}{*}{$\begin{array}{c}\rm Measured \vspace{-1pt} \\ \rm upper~limit \vspace{-1pt} \\
\rm at~95\%\,CL~\mbox{\cite{pdg}}\end{array}$} & \multicolumn{4}{c|}{Prediction maximum [or range]} \\
\cline{3-6} & & \multicolumn{2}{c|}{NO} & \multicolumn{2}{c|}{IO} \\ \cline{3-6}
& & \scriptsize 0.6 TeV & \scriptsize 1 TeV & \scriptsize0.6 TeV & \scriptsize 1 TeV
\\ \hline\hline
$Z\to e^\pm\mu^\mp\vphantom{|^{\int^|}}$   & $7.5\times10^{-7}$  & $8.3\times10^{-10}$ &
$1.8\times10^{-11}$ & ~$8.3\times10^{-10}$~ & ~$1.8\times10^{-11}$~ \\
$Z\to e^\pm\tau^\mp\vphantom{|^{\int^|}}$    & $9.8\times10^{-6}$ & $3.2\times10^{-6}$ &
$7.0\times10^{-8}$ & \,$4.7\times10^{-7}$\, & \,$1.0\times10^{-8}$\, \\
\,$Z\to\mu^\pm\tau^\mp\vphantom{|^{\int^|}}$\, & $1.2\times10^{-5}$ &
~$[0.8,8.5]\times\!10^{-7}$~ & ~$[0.2,1.9]\times10^{-8}$~ &
$8.8\times10^{-7}$ & $1.9\times10^{-8}\vphantom{|_{|_O}^{}}$
\\ \hline
\end{tabular} \smallskip
\caption{The maximum predictions of the branching fractions of \,$Z\to e\mu,e\tau,\mu\tau$\,
due to loop contributions of the $Z'$ with mass \,$m_{Z'}^{}=0.6$ and 1~TeV,\, compared to
the experimental limits.  The lower end of a prediction is also displayed if exceeding one per
mill of its upper end.} \label{Z->ll'}
\end{table}

The flavor-violating $Z$-boson decays \,$Z\to\ell\bar\ell'$\, also are not yet observed, but
there have been searches for them resulting in the limits quoted in Table\,\,\ref{Z->ll'}.
These processes can happen here because of flavor-violating $Z'$-loop modifications
to the $Z\ell\bar\ell$ vertex and leptonic self-energy
diagrams~\cite{Chiang:2013aha,Carone:1994aa}.
From the decay amplitude
\,${\cal M}_{Z\to\ell\bar\ell'}=\bar u_\ell\,\slashed\varepsilon_Z
(L_{\ell\ell'}P_L + R_{\ell\ell'}P_R) v_{\bar\ell'}$,\,
one arrives at the rate
\begin{eqnarray}
\Gamma_{Z\to\ell\bar\ell'}^{} &=& \frac{|\textbf{\textsl p}_\ell^{}|}{12\pi m_Z^2}
\Bigg\{ \big(|L_{\ell\ell'}^{}|^2+|R_{\ell\ell'}^{}|^2\big)
\Bigg[m_Z^2-\frac{m_\ell^2+m_{\ell'}^2}{2}-\frac{(m_\ell^2-m_{\ell'}^2)^2}{2 m_Z^2}\Bigg]
\nonumber \\ && \hspace{9ex} +\;
6\, {\rm Re}\big(L_{\ell\ell'}^*R_{\ell\ell'}^{}\big)\, m_{\ell'}^{}m_\ell^{} \Bigg\} \,,
\end{eqnarray}
where $\textbf{\textsl p}_\ell^{}$ is the three-momentum of $\ell$  in the $Z$ rest-frame.
Including the SM and $Z'$ contributions, one has
\begin{eqnarray} \label{LfRf}
L_{\ell\ell'}^{} \,=\, \delta_{\ell\ell'\,}^{}g_L^{\textsc{sm}} + L_{\ell\ell'}^{Z'} \,, ~~~~~
R_{\ell\ell'}^{} \,=\, \delta_{\ell\ell'\,}^{}g_R^{\textsc{sm}} \,,
\end{eqnarray}
where
\,$g_L^{\textsc{sm}}=g\big(2s_{\rm w}^2-1\big)/(2c_{\rm w}^{})$\, and
\,$g_R^{\textsc{sm}}=gs_{\rm w}^2/c_{\rm w}^{}$\, are the SM contributions at tree level, with
$g$ being the weak coupling constant, \,$c_{\rm w}^{}=\sqrt{1-s_{\rm w}^2}$,\, and $s_{\rm w}^2$
the squared sine of the Weinberg angle.
In terms of the elements of $\Delta_\ell$, the $Z'$ part is given by~\cite{Chiang:2013aha}
\begin{eqnarray} \label{Z2ll'Z'}
L_{\ell_k\ell_l}^{Z'} &=& \frac{-{\cal F}(\varrho)}{16\pi^2}\,
\raisebox{2pt}{\footnotesize$\displaystyle\sum_o$}\,(\Delta_\ell)_{ko}^{}(\Delta_\ell)_{ol}^{} \,,
~~~~~ \varrho \,=\, \frac{m_{Z'}^2}{m_Z^2} \,,
\nonumber \\
{\cal F}(\varrho) &=& \frac{7}{2} + 2\varrho
+ 2(1+\varrho)^2\,{\rm Li}_2^{}\biggl(-\frac{1}{\varrho}\biggr) + (\ln\varrho+i\pi)
\biggl[ 3+2\varrho+ 2(1+\varrho)^2\,\ln\frac{\varrho}{1+\varrho} \biggr] \,. ~~~
\end{eqnarray}
Numerically, we have checked that for \,$\ell'=\ell$\, the $Z'$ benchmark points extracted above
produce effects on the $Z$-pole observables that are well within the 2$\sigma$ ranges of their
data~\cite{pdg}, as long as~\,$m_{Z'}^{}\,\mbox{\footnotesize$\gtrsim$}\,0.5$\,TeV.\,
At the same time, for \,$\ell'\neq\ell$\, the $Z'$ contributions to \,$Z\to\ell\bar\ell'$\,
may be observable in the not-too-distant future.
In Table\,\,\ref{Z->ll'}, from our benchmarks we present predictions for the branching
fractions of these LFV decays for \,$m_{Z'}^{}=0.6$ and 1~TeV.\,
These examples illustrate that \,$Z\to e\mu$\, is unlikely to be detectable soon.
Nevertheless, the numbers for \,$Z\to e\tau$\, and \,$Z\to\mu\tau$\, can be less than 20 times
below the corresponding experimental bounds, but are mostly of order $10^{-8}$-$10^{-7}$.
Thus, one or two of these predictions may already be within the reach of the upcoming
High-Luminosity LHC (HL-LHC), which is expected to improve upon the present limits by factors
of a few with a luminosity of $200\,{\rm fb}^{-1}$~\cite{Davidson:2012wn}.
More powerful $Z$ factories are therefore necessary to test more of the predictions in this table.
For instance, the GigaZ option of a future $e^+e^-$ collider can produce at least $10^9$ $Z$s and
be sensitive to LFV $Z$ decays at the $10^{-9}$ level \cite{AguilarSaavedra:2001rg,gigaz}.
Much more promising is the FCC-$ee$, which can achieve sensitivity up to $\mathcal O(10^{-13})$
with $10^{13}$ $Z$s~\cite{Blondel:2014bra}.

\section{Conclusions\label{sec:summary}}

Inspired by the recent hint of lepton flavor nonuniversality in the \,$B\to K^*\mu\bar\mu$\, and
\,$K^*e\bar e$\, decays, along with several other anomalies observed earlier in
\,$b\to s\ell\bar\ell$ transitions, we have studied within the minimal flavor violation
framework whether the parameter space preferred by such data can be consistent with
a wider class of observables.
Restricting ourselves to new physics operators up to dimension 6, we have shown that the new
interactions are chiral and feature a specific relation for the Wilson coefficients in
the effective Hamiltonian for \,$b\to s\ell\bar\ell$\, decays:
\,$C_9^{\ell, {\rm NP}} = - C_{10}^{\ell, {\rm NP}}$.\,
With the hierarchy in quark Yukawa couplings and the assumption of ${\cal O}(1)$ neutrino Yukawa
couplings, we have found that only the couplings involving $\Delta_q$ and $\Delta_\ell$, defined
in Eq.~\eqref{DqDl}, can induce flavor-violating interactions.

We have also considered a scenario where the new physics effects on the \,$b\to s\ell\bar\ell$\,
decays are caused by a $Z'$ gauge boson with nonuniversal couplings to SM fermions.
Moreover, we require these couplings to respect the MFV principle, parametrizing them with
the elements of $\Delta_{q,\ell}$.  The $Z'$ boson is assumed in particular to have no
flavor-conserving coupling to the electron.  These new interactions lead to dimension-6 operators
with flavor violation that are constrained by the limits or measurements of various observables.
Out of them, we find that the $B$-$\bar B$ mixing data are very consequential and the empirical
bounds for \,$\mu\to e\gamma$\, and \,$\tau\to3\mu$\, often play major roles in further
constraining the parameter space in the model.

Through numerical scans of the coefficients in $\Delta_{q,\ell}$ and the neutrino oscillation
parameters for both the normal and inverted orderings of the light neutrinos' masses,
we have obtained sampling benchmark points for our $Z'$ scenario that are compatible with
the different constraints.
The viable parameter space depends highly on the structure of the ${\sf A}_\ell$ matrix constructed
from the right-handed neutrinos' Yukawa couplings and on the light neutrinos' mass ordering.
With the simplest form of ${\sf A}_\ell$, only the NO case possesses viable parameter space,
albeit marginally.
Adopting a less simple choice of ${\sf A}_\ell$ with extra complex phases, we demonstrate that both
the NO and IO scenarios have good amounts of allowed parameter space, with the IO case being preferred,
and subsequently we predict a number of observables.  Our predictions concern mostly
lepton-flavor-violating modes in charged-lepton decays, $b$-meson and kaon decays, and $Z$-boson
decays, but we also evaluate the $Z'$ impact on \,$b\to s\tau\bar\tau$\, and rare meson decays
involving neutrinos.
The upper bounds of our estimates for the rates of some of these processes can be further probed by
searches or measurements in the near future.

\acknowledgments

This work was supported in part by the Ministry of Education (MOE) Academic Excellence Program
(Grant No. 105R891505) and National Center for Theoretical Sciences (NCTS) of the Republic of China (ROC).
C.W.C. was also supported in part by the Ministry of Science and Technology (MOST) of ROC (Grant
No.~MOST~104-2628-M-002-014-MY4).
X.G.H. was also supported in part by MOST of ROC (Grant No. MOST104-2112-M-002-015-MY3) and in part
by the National Science Foundation of China (NSFC) (Grant Nos. 11175115, 11575111 and 11735010),
Key Laboratory for Particle Physics,
Astrophysics and Cosmology, Ministry of Education, and Shanghai Key Laboratory for Particle
Physics and Cosmology (SKLPPC) (Grant No. 15DZ2272100) of the People's Republic of
China (PRC).

\appendix

\section{Extra constraints on \boldmath$Z'$ couplings\label{2q2l}}

At tree level, the $Z'$ interactions in Eq.\,(\ref{LZ'}) contribute to \,$\mu\to e$\, conversion
in nuclei via the operator ${\cal O}^{2q2\ell}$ in Eq.\,(\ref{Lmfv}).
To calculate the branching fraction ${\cal B}(\mu {\cal N}\to e{\cal N})$ of \,$\mu\to e$\,
conversion in nucleus $\cal N$, we employ the pertinent formulas provided in
Ref.\,\cite{Kitano:2002mt}.
Thus, we arrive at
\begin{eqnarray} \label{Bm2e}
{\cal B}(\mu {\cal N}\to e{\cal N}) &=& \frac{m_{\mu\,}^5
\big|\big(2 g_{uue\mu}^{}+g_{dde\mu}^{}\big)V_{\cal N}^p
+ \big(g_{uue\mu}^{}+2 g_{dde\mu}^{}\big)V_{\cal N}^n\big|\raisebox{2pt}{$^2$}}
{\omega_{\rm capt}^{\cal N}} \,,
\\ \nonumber \\ \label{gqqem}
g_{uue\mu}^{} &=& \frac{\big(V_{\textsc{ckm}}^\dagger
\Delta_q^{}V_{\textsc{ckm}}^{}\big)_{11}(\Delta_\ell)_{12}^{}}{m_{Z'}^2}
\,=\, \frac{\big(\zeta_0^{}+\zeta_1^{~}y_u^2+\zeta_2^{}y_u^4\big)
(\Delta_\ell)_{12}^{}}{m_{Z'}^2} \,,
\nonumber \\
g_{dde\mu}^{} &=& \frac{(\Delta_q)_{11}^{}(\Delta_\ell)_{12}^{}}{m_{Z'}^2}
\,=\, \frac{\big[\zeta_0^{}+|V_{td}|^2\big(\zeta_1^{~}y_t^2+\zeta_2^{~}y_t^4\big)\big]
(\Delta_\ell)_{12}^{}}{m_{Z'}^2} \,,
\end{eqnarray}
where $V_{\cal N}^{p(n)}$ is an overlap integral for the protons (neutrons) in $\cal N$ and
$\omega_{\rm capt}^{\cal N}$ the rate of muon capture in ${\cal N}$.
Based on the data on \,$\mu\to e$\, transition in nuclei~\cite{pdg} and the corresponding
$V_{\cal N}^{p(n)}$ and $\omega_{\rm capt}^{\cal N}$ values~\cite{Kitano:2002mt}, we find
the gold limit
\,${\cal B}(\mu{\rm Au}\to e{\rm Au})_{\rm exp}^{}<7.0\times10^{-13}$ at 90\% CL~\cite{pdg}
to supply the strictest restraint.
Using \,$V_{\rm Au}^{p(n)}=0.0974\,(0.146)$ and \,$\omega_{\rm capt}^{\rm Au}=13.07\times10^6$/s\,
\cite{Kitano:2002mt}, we then extract
\begin{eqnarray} \label{xguxgd}
\big|g_{uue\mu}^{}+1.14\,g_{dde\mu}^{}\big| \,<\, \frac{2.0\times10^{-6}}{\rm TeV^2} \,.
\end{eqnarray}
Since our benchmark points from the permitted parameter space in the NO (IO) case
yield the bound \,$\big|(\Delta_\ell)_{12}^{}\big|/m_{Z'}^{}<0.065\,(0.067)$/TeV,\, while
\,$\big|\zeta_1^{~}y_t^2+\zeta_2^{~}y_t^4\big|/m_{Z'}^{}<0.13$/TeV\, from Eq.\,(\ref{zyt}),
and \,$y_u^2\sim10^{-10}$\, and \,$|V_{td}|^2\sim7\times10^{-5}$\, from quark data \cite{pdg},
it is evident that by choosing
\,$\big|\zeta_0^{}\big|/m_{Z'}^{}\;\mbox{\footnotesize$\lesssim$}\;8\!\times\!10^{-6}$/TeV\,
in Eq.\,(\ref{gqqem}) we can make the $Z'$ contributions compatible with the condition in
Eq.\,(\ref{xguxgd}).

The recent LHC measurements on \,$pp\to\mu^+\mu^-$ \cite{ATLAS:2017wce} translate into
restrictions on potential NP affecting the partonic reactions \,$\bar qq\to\mu^+\mu^-$.\,
The relevant $Z'$ couplings are
\begin{eqnarray} \label{gqqmm}
g_{uu\mu\mu}^{} &\simeq& \frac{\zeta_0^{}\,(\Delta_\ell)_{22}^{}}{m_{Z'}^2} \,,\, \hspace{7em}
g_{dd\mu\mu}^{} \,=\, \frac{\big[\zeta_0^{}+|V_{td}|^2\big(\zeta_1^{~}y_t^2+\zeta_2^{~}y_t^4\big)\big]
(\Delta_\ell)_{22}^{}}{m_{Z'}^2} \,,
\nonumber \\
g_{cc\mu\mu}^{} &\simeq& \frac{\big(\zeta_0^{}+\zeta_1^{~}y_c^2\big)
(\Delta_\ell)_{22}^{}}{m_{Z'}^2} \,, \hspace{8ex}
g_{ss\mu\mu}^{} \,=\, \frac{\big[\zeta_0^{}+|V_{ts}|^2\big(\zeta_1^{~}y_t^2+\zeta_2^{~}y_t^4\big)\big]
(\Delta_\ell)_{22}^{}}{m_{Z'}^2} \,,
\nonumber \\
g_{bb\mu\mu}^{} &=& \frac{\big[\zeta_0^{}+|V_{tb}|^2
\big(\zeta_1^{~}y_t^2+\zeta_2^{~}y_t^4\big)\big](\Delta_\ell)_{22}^{}}{m_{Z'}^2} \,.
\end{eqnarray}
From the aforementioned benchmarks, we get
\,$\big|(\Delta_\ell)_{22}^{}\big|/m_{Z'}^{}<0.14\,(0.26)$/TeV\, in the NO (IO) case.
Then, with \,$|V_{ts}|^2\sim0.0016$, \,$|V_{tb}|^2\sim1$,\, and
\,$y_c^2\sim2\!\times\!10^{-5}$ \cite{pdg}, as well as the other parameter values specified in
the preceding paragraph, we can derive, in units of TeV$^{-2}$,
\begin{eqnarray} \label{qqmm}
\big|g_{uu\mu\mu}^{}\big| &\mbox{\footnotesize$\lesssim$}& 2.1\times10^{-6} \,, ~~~
\big|g_{dd\mu\mu}^{}\big| \;\mbox{\footnotesize$\lesssim$}\; 4.4\times10^{-6} \,, ~~~
\big|g_{cc\mu\mu}^{}\big| \;\mbox{\footnotesize$\lesssim$}\; 7.3\times10^{-6} \,, ~~~
\big|g_{ss\mu\mu}^{}\big| \;\mbox{\footnotesize$\lesssim$}\; 5.6\times10^{-5} \,,
\nonumber \\
\big|g_{bb\mu\mu}^{}\big| &\mbox{\footnotesize$\lesssim$}& 0.034 \,,
\end{eqnarray}
after additionally selecting \,$\zeta_1^{}\sim m_{Z'}^{}$/TeV\, for $g_{cc\mu\mu}^{}$.
Most of these numbers are at least three orders of magnitude below their respective bounds
inferred in Ref.\,\,\cite{Greljo:2017vvb} from the \,$pp\to\mu^+\mu^-$ data
\cite{ATLAS:2017wce}, except
\,$-0.38\;\mbox{\footnotesize$\lesssim$}\;g_{bb\mu\mu}^{\rm exp}\,{\rm TeV}^2
\;\mbox{\footnotesize$\lesssim$}\;0.46$,\,
which is still more than an order of magnitude above its $Z'$ counterpart in Eq.\,(\ref{qqmm}).

\end{document}